\documentclass[a4paper,11pt]{article}
\usepackage{amsmath,amssymb,bbm}
\usepackage{ascmac}
\usepackage{cancel}
\usepackage{graphicx}
\usepackage{hyperref}
\hypersetup{
pdftitle={Worldsheet description of exotic five-brane with two gauged isometries},
pdfauthor={Tetsuji Kimura and Shin Sasaki},
colorlinks={true},
linkcolor={blue},
urlcolor={blue},
filecolor={blue},
citecolor={blue}
}
\makeatletter

 \@addtoreset{equation}{section}
\makeatother

\newcounter{Enumerate}

\DeclareFontFamily{U}{rsf}{}
\DeclareFontShape{U}{rsf}{m}{n}{
  <5> <6> rsfs5 <7> <8> <9> rsfs7 <10-> rsfs10}{}
\DeclareMathAlphabet\Scr{U}{rsf}{m}{n}

\usepackage[mathscr]{eucal}

\newcommand{\del}{\partial}
\newcommand{\half}{\frac{1}{2}}

\newcommand{\ls}{\ \ \ \ \ }
\newcommand{\wt}{\widetilde}
\newcommand{\wh}{\widehat}
\newcommand{\ve}{\varepsilon}
\newcommand{\ol}{\overline}

\newcommand{\bsubeq}{\begin{subequations}}
\newcommand{\esubeq}{\end{subequations}}

\newcommand{\noi}{\noindent}

\newcommand{\nn}{\nonumber}

\newcommand{\A}{\mathscr{A}}
\newcommand{\B}{\mathscr{B}}
\newcommand{\C}{\mathscr{C}}

\newcommand{\N}{\mathcal{N}}
\renewcommand{\d}{{\rm d}}
\newcommand{\e}{{\rm e}}

\renewcommand{\i}{{\rm i}}

\newcommand{\slb}{\scalebox}

\def\+{{+\!\!\!+}}

\begin{document}
\allowdisplaybreaks{

\thispagestyle{empty}


\begin{flushright}
RUP-13-13
\end{flushright}

\vspace{35mm}

\noi
\slb{2.15}{Worldsheet Description of Exotic Five-brane}

\vspace{5mm}

\noi
\slb{2.15}{with Two Gauged Isometries}

\vspace{15mm}

\slb{1.2}{Tetsuji {\sc Kimura}$^{\,a}$} \ and \ \slb{1.2}{Shin {\sc Sasaki}$^{\,b}$}

\slb{.85}
{\renewcommand{\arraystretch}{1.2}
\begin{tabular}{rl}
$a$ & {\sl
Department of Physics \& Research Center for Mathematical Physics,
Rikkyo University} 
\\
& {\sl 
Tokyo 171-8501, JAPAN
}
\\
& {\tt tetsuji \_at\_ rikkyo.ac.jp}
\end{tabular}
}

\slb{.85}
{\renewcommand{\arraystretch}{1.2}
\begin{tabular}{rl}
$b$ & {\sl
Department of Physics,
Kitasato University}
\\
& {\sl 
Sagamihara 252-0373, JAPAN}
\\
& {\tt shin-s \_at\_ kitasato-u.ac.jp}
\end{tabular}
}

\vspace{15mm}


\begin{abstract}
We study the string worldsheet description of the background geometry of the exotic $5^2_2$-brane where two isometries are gauged.
This is an extension of the gauged linear sigma model (GLSM) for the exotic $5^2_2$-brane with a single gauged isometry. 
The new GLSM with two gauged isometries has only ${\mathcal N}=(2,2)$ supersymmetry rather than ${\mathcal N}=(4,4)$ supersymmetry of the original GLSM.
This is caused by a conflict between two different $SU(2)_R$ associated with the two gauge symmetries.
However, if we take a certain limit, we can find the genuine string sigma model of the background geometry of the exotic $5^2_2$-brane
with ${\mathcal N}=(4,4)$ supersymmetry.
We also investigate the worldsheet instanton corrections to the background geometry of the exotic $5^2_2$-brane.
The worldsheet instanton corrections to the string sigma model can be traced in terms of the two gauge fields in the new GLSM.
This new GLSM gives rise to a different feature of the quantum corrections from the one in the GLSM with the single gauged isometry.
\end{abstract}

\newpage
\section{Introduction}

What is the fundamental unit of matters and spacetime in Nature?
We have not obtained the answer of this question yet,
because the quantum theory of {\it everything} has not been found.
Even though the Standard Model for matter fields and their interactions has been established in our energy level, 
we have not known more fundamental aspects of it in much higher energy scale.
String theory is expected as a candidate to describe the quantum feature of gravitational force as well as the origin of matter fields and gauge interactions.
However, it is still very hard to understand the quantum aspects of string theory.

D-branes, the extended objects in string theory, play a central role to understand the nonperturbative feature of string theory.
Indeed they contribute to the quantum aspects of black hole physics.
However, one suspects that counting D-branes' dynamics is still not enough to describe the quantum behavior of the black hole physics completely.
Exploring the black holes physics, 
one has recognized that ``exotic branes'' should be involved
\cite{Elitzur:1997zn, Hull:1997jc, Blau:1997du, Obers:1997kk}.
Exotic branes are non-standard objects 
because the metric of their background geometry 
is not a single-valued function.
The lack of the single-valuedness is caused by the mixing of the diffeomorphism  and the duality transformations in string theory.
The exotic branes have already been argued in 
\cite{Elitzur:1997zn, Hull:1997jc, Blau:1997du, Obers:1997kk, Obers:1998fb},
and are exhaustively discussed in \cite{deBoer:2010ud, deBoer:2012ma} again.

Conventionally, the spacetime metric is single-valued if the spacetime is probed by a point particle.
In this case the spacetime geometry is represented by the Riemann(-Cartan) geometry.
However, when the spacetime is probed by a string,
the description of the spacetime geometry should be extended, i.e.,
the structure of ``winding'' should be involved.
In other words,
the spacetime geometry is reformulated in terms of the metric $G$ and the NS-NS B-field $B$.
Once information of the winding is involved in the spacetime geometry,
it contains the structure of the string {\rm T}-duality in a very natural way.
Such a geometry is described by 
``generalized geometry'' \cite{Hitchin:2004ut}.
Soon after, 
string theorists applied the generalized geometry to flux compactification scenarios (for a comprehensive review, see \cite{Grana:2005jc}).
The generalized geometry exhibits not only the conventional Riemann(-Cartan) geometry, 
but also ``nongeometry'' which is not captured by the conventional metric only.
Historically, however,
the conventional geometry given only by the metric has been exhaustively studied even in string theory.
The generalized geometry, i.e., the geometry by the metric and the B-field, has not been argued seriously.
This is because people think
that the generalized geometry is a bit far from the conventional geometry which is familiar with the general theory of relativity.
We now encounter difficulties 
of the analysis of the black hole quantum mechanics, as far as we concern it only in terms of the conventional geometry.
Thus it is the time to study the generalized geometry in order to evaluate the quantum aspects of the black hole physics completely.
Indeed, the exotic brane is the typical object of this study.

In this work we focus on the exotic $5^2_2$-brane.
This is a typical example of exotic branes.
We begin with the background geometry of a single H-monopole, i.e.,
a single NS5-brane smeared along a compact $S^1$-circle.
Performing the {\rm T}-duality transformation 
followed by the Buscher rule \cite{Buscher:1987sk}, 
we obtain the background geometry of a single Kaluza-Klein (KK) monopole.
This geometry is described as the (single-centered) Taub-NUT space.
Now we further compactify one of the three directions of the Taub-NUT space.
The KK-monopole is reduced to the five-brane of codimension two,
whilst the original KK-monopole and the H-monopole are of codimension three.
(We often refer to five-branes of codimension two as defect five-branes \cite{Bergshoeff:2011se}.)
If we take the {\rm T}-duality transformation along the second compact direction,
we find a new object of codimension two.
This is the exotic $5^2_2$-brane.
The background geometry is regarded as a concrete example of 
{\rm T}-folds \cite{Hull:2004in} with (globally) nongeometric structure.


Since we investigate the exotic brane beyond the supergravity descriptions,
we are interested in the string worldsheet description:
\begin{align}
\Scr{L}_{\text{string}}
\ &= \ 
- \half G_{IJ} \, g^{mn} \, \del_m X^I \del_n X^J
+ \half B_{IJ} \, \ve^{mn} \, \del_m X^I \del_n X^J
\nn \\
\ & \ \ \ \ 
+ \frac{\i}{2} G_{IJ} \, \Omega^I_- \nabla_+ \Omega^J_-
+ \frac{\i}{2} G_{IJ} \, \Omega^I_+ \nabla_- \Omega^J_+
+ \frac{1}{4} R_{IJKL} \, \Omega^I_+ \Omega^J_+ \Omega^K_- \Omega^L_-
\, . \label{string-NLSM}
\end{align}
Here $G_{IJ}$ is the spacetime metric and $B_{IJ}$ is the NS-NS B-field
in ten dimensions,
whilst $g_{mn}$ is the worldsheet metric 
and $\ve^{mn}$ is the two-dimensional Levi-Civita tensor normalized as 
$\ve^{01} = + 1 = - \ve_{01}$.
The real fermions $\Omega^I_{\pm}$ are superpartners of the string coordinate fields $X^I$.
The covariant derivatives $\nabla_{\pm}$ carry the affine connection of the target space geometry. 
The coefficient $R_{IJKL}$ of the four-fermions term denotes the Riemann curvature of the target space.
Note that the indices $I, J$ run from 6 to 9, 
which represent the transverse directions of the five-brane.
The indices $m,n$ are the ones of the worldsheet coordinates.
Since we consider the spacetime geometry in the string frame,
the longitudinal directions $012345$ of the five-brane
are flat. 
Thus the coordinate fields of these directions are decoupled from the string worldsheet sigma model 
(for the spacetime indices, see Table \ref{ST-WS} in appendix \ref{app-SF}).

It is worth extending the string worldsheet theory to a gauge theory of specific type called the gauged linear sigma model (GLSM) \cite{Witten:1993yc}.
In the case of $\N=(2,2)$ GLSM,
its low energy effective theory in the IR limit can be described as a nonlinear sigma model (NLSM) or a Landau-Ginzburg theory, 
both are useful field theories to describe string worldsheet dynamics.
The string worldsheet instanton corrections can be traced by the soliton configurations of the gauge theory, 
which can be interpreted as the quantum deformations of the K\"{a}hler moduli.
An $\N=(4,4)$ GLSM is also quite useful to study NS5-branes 
and Kaluza-Klein (KK) monopoles \cite{Tong:2002rq, Harvey:2005ab, Okuyama:2005gx}, since
the quantum aspects of the five-branes, i.e., the worldsheet instanton corrections, can be traced by the solitonic description of the $\N=(4,4)$ gauge theory.
We applied the $\N=(4,4)$ GLSM to the classical description of the exotic $5^2_2$-brane \cite{Kimura:2013fda}, and studied the worldsheet instanton corrections of the $5^2_2$-brane \cite{Kimura:2013zva}.

In this paper, we develop our previous works \cite{Kimura:2013fda, Kimura:2013zva} in order to investigate other quantum corrections to the exotic $5^2_2$-brane.
We notice that the background geometry of the exotic $5^2_2$-brane has two independent isometries along the transverse directions.
In the previous works we considered the sigma model in which only one of the two isometries are gauged.
It is natural to think of an extension of the sigma model to the one where both of the two isometries are gauged.
Actually we find a new insight of the sigma model,
even in the lack of the complete understanding of it.

The organization of this paper is as follows: 
In section \ref{review} we review the $\N=(4,4)$ GLSMs for multiple H-monopoles and for multiple KK-monopoles.
Performing the duality transformations, 
we obtain the GLSM for the exotic $5^2_2$-brane.
In the IR limit we find the NLSM whose target space is the background geometry of the exotic $5^2_2$-brane.
In section \ref{sect-2-gauged} we construct a GLSM with two gauged isometries.
This is an extension of the GLSM for the exotic $5^2_2$-brane \cite{Kimura:2013fda}. 
We refer to this model as the Remodeled GLSM.
First we mention the supersymmetry which this model has.
Next we investigate the classical feature in the IR limit.
Third we argue the restoration of the supersymmetry which is consistent with the background geometry of the $5^2_2$-brane.
In section \ref{Instantons} we study the quantum corrections to the Remodeled GLSM. 
We can argue the worldsheet instanton corrections from the vortex configurations of the gauge fields.
However, we find that the vortex corrections from the second gauge field do not contribute to the background geometry of the exotic $5^2_2$-brane even if we choose the most reliable parameter regime.
Section \ref{summary} is devoted to summary and discussions. 
In appendix \ref{app-SF} we exhibit the conventions of two-dimensional $\N=(2,2)$ supersymmetry.
In appendix \ref{app-KKM2-H2} we briefly discuss the Remodeled GLSMs for other defect five-branes.

\section{Review of GLSMs for five-branes}
\label{review}

In this section we briefly review a construction of the worldsheet sigma model whose target space is the background geometry of the exotic $5^2_2$-brane \cite{Kimura:2013fda}.
As we mentioned before, the background geometry of the exotic $5^2_2$-brane is obtained via two transverse {\rm T}-duality transformations on an NS5-brane.
Fortunately, we have already known the GLSM for the H-monopole, which is the NS5-brane smeared along a compact $S^1$-circle \cite{Tong:2002rq}.
Performing the duality transformations \cite{Rocek:1991ps} to the GLSM for the H-monopole,
and taking the IR limit,
we find that the supersymmetric NLSM whose target space is the geometry of the exotic $5^2_2$-brane.

We begin with the following $\N=(4,4)$ supersymmetric GLSM \cite{Tong:2002rq}:
\begin{align}
\Scr{L}_{\text{H}} \ &= \ 
\sum_{a=1}^k \int \d^4 \theta \, \Big\{ 
\frac{1}{e_a^2} 
\Big( - \ol{\Sigma}{}_a \Sigma_a + \ol{\Phi}{}_a \Phi_a \Big)
+ \ol{Q}{}_a \, \e^{-2 V_a} Q_a
+ \ol{\wt{Q}}{}_a \, \e^{+2 V_a} \wt{Q}_a
\Big\}
\nn \\
\ & \ \ \ \ 
+ \int \d^4 \theta \, 
\frac{1}{g^2} 
\Big( - \ol{\Theta} \Theta + \ol{\Psi} \Psi \Big)
+ \sum_{a=1}^k \Big\{
\sqrt{2} \int \d^2 \theta \, \Big( \wt{Q}_a \Phi_a Q_a + (s_a - \Psi) \Phi_a
\Big)
+ \text{(h.c.)} 
\Big\}
\nn \\
\ & \ \ \ \ 
+ \sum_{a=1}^k \Big\{ 
\sqrt{2} \int \d^2 \wt{\theta} \, \big( t_a - \Theta \big) \Sigma_a
+ \text{(h.c.)}
\Big\}
\, . \label{GLSM-HM}
\end{align}
An $\N=(4,4)$ abelian vector multiplet is denoted by an $\N=(2,2)$ abelian vector superfield $V_a$ (or a twisted chiral superfield $\Sigma_a = \frac{1}{\sqrt{2}} \ol{D}{}_+ D_- V_a$) and an $\N=(2,2)$ chiral superfield $\Phi_a$.
An $\N=(4,4)$ charged hypermultiplet is given by a set of $\N=(2,2)$ chiral superfields $(Q_a, \wt{Q}_a)$, 
where $Q_a$ ($\wt{Q}_a$) has charge $-1$ ($+1$) under the $U(1)$ gauge transformation.
The pair $(\Psi, \Theta)$ belongs to an $\N=(4,4)$ neutral hypermultiplet,
where $\Psi$ is an $\N=(2,2)$ chiral superfield, 
whilst $\Theta$ is an $\N=(2,2)$ twisted chiral superfield\footnote{See appendix \ref{app-SF} for the expansion rule of $\N=(2,2)$ superfields in terms of component fields (see also \cite{Kimura:2013fda}).}.
Each vector multiplet $(V_a, \Phi_a)$ has a set of complex-valued Fayet-Iliopoulos (FI) parameters $(s_a, t_a)$.
The gauge coupling constant $e_a$ has mass dimension one, 
while the sigma model coupling constant $g$ is dimensionless.
The model (\ref{GLSM-HM}) becomes an $\N=(4,4)$ supersymmetric theory 
if we impose suitable representations of $SU(2)_R$ symmetry on the component fields of the above superfields in a consistent way \cite{Tong:2002rq}. 
An explicit assignment of the representations of the $SU(2)_R$ symmetry is discussed in \cite{Harvey:2005ab}.

In the IR limit of the gauge theory (\ref{GLSM-HM}), 
all the gauge multiplets are integrated out
and all the charged hypermultiplets are solved in terms of the neutral hypermultiplet.
The scalar components of the $\N=(4,4)$ neutral hypermultiplet denote the coordinates of the target space ${\mathbb R}^3 \times S^1$.
The low energy effective theory can be interpreted as the string worldsheet sigma model whose target space denotes the background geometry of multi-centered H-monopoles of codimension three and the NS-NS B-field\footnote{We only focus on the transverse four directions of five-branes in the string frame.}. 
The detailed derivation can be seen in \cite{Tong:2002rq, Harvey:2005ab, Okuyama:2005gx}.
A benefit of the gauge theory (\ref{GLSM-HM}) is that the string worldsheet instanton corrections can be computed by vortex corrections in the gauge theory framework \cite{Tong:2002rq}.
Indeed, 
the alignment of the H-monopoles along the compact $S^1$-circle are affected by the vortex corrections.
This implies that the string worldsheet instanton corrections yield the KK-momentum corrections to the background geometry of the H-monopoles.

Applying the duality transformations \cite{Rocek:1991ps} to the twisted-chiral superfield $\Theta$ in (\ref{GLSM-HM}), 
we obtain the following gauge theory:
\begin{align}
\Scr{L}_{\text{KK}} \ &= \ 
\sum_{a=1}^k \int \d^4 \theta \, \Big\{ 
\frac{1}{e_a^2} 
\Big( - \ol{\Sigma}{}_a \Sigma_a + \ol{\Phi}{}_a \Phi_a \Big)
+ \ol{Q}{}_a \, \e^{-2 V_a} Q_a
+ \ol{\wt{Q}}{}_a \, \e^{+2 V_a} \wt{Q}_a
\Big\}
\nn \\
\ & \ \ \ \
+ \int \d^4 \theta \, \Big\{
\frac{1}{g^2} \ol{\Psi} \Psi
+ \frac{g^2}{2} 
\Big( \Gamma + \ol{\Gamma} + 2 \sum_{a=1}^k V_a \Big)^2
\Big\}
\nn \\
\ & \ \ \ \ 
+ \sum_{a=1}^k \Big\{
\sqrt{2} \int \d^2 \theta \, \Big( \wt{Q}_a \Phi_a Q_a + (s_a - \Psi) \Phi_a
\Big)
+ \text{(h.c.)} 
\Big\}
\nn \\
\ & \ \ \ \ 
+ \sum_{a=1}^k \Big\{ 
\sqrt{2} \int \d^2 \wt{\theta} \, t_a \Sigma_a
+ \text{(h.c.)}
\Big\}
- \sqrt{2} \, \ve^{mn} \sum_{a=1}^k \del_m (\vartheta A_{a,n})
\, . \label{GLSM-KKM}
\end{align} 
This is also an $\N=(4,4)$ supersymmetric theory.
Now $\Gamma$ is an $\N=(2,2)$ chiral superfield dualized from the twisted chiral superfield $\Theta$ under the relation
\begin{align}
\Theta + \ol{\Theta}
+ 2 g^2 \sum_{a=1}^k V_a
\ &= \ 
- g^2 (\Gamma + \ol{\Gamma}) 
\, . \label{Theta2Gamma}
\end{align} 
In the IR limit, 
the gauge theory (\ref{GLSM-KKM}) is reduced to the string worldsheet sigma model of multi-centered KK-monopoles of codimension three, since 
the target space of the sigma model is the multi-centered Taub-NUT space.
The detailed derivation can be also seen in \cite{Tong:2002rq, Harvey:2005ab, Okuyama:2005gx}.
In this model the vortex corrections by the gauge fields are also interpreted as the string worldsheet instanton corrections.
Notice that the geometrical interpretation of the instanton corrections is now the winding corrections to the Taub-NUT space, 
rather than the KK-momentum corrections \cite{Harvey:2005ab}. 
This is consistent with ``T-duality'' of the worldsheet instanton corrections to the background geometry of the H-monopoles mentioned above.

If we further apply the duality transformation  
to the chiral superfield $\Psi$ in (\ref{GLSM-KKM}),
we obtain the following new GLSM:
\begin{align}
\Scr{L}_{\text{E}} \ &= \ 
\sum_{a=1}^k \int \d^4 \theta \, \Big\{ 
\frac{1}{e_a^2} 
\Big( - \ol{\Sigma}{}_a \Sigma_a + \ol{\Phi}{}_a \Phi_a \Big)
+ \ol{Q}{}_a \, \e^{-2 V_a} Q_a
+ \ol{\wt{Q}}{}_a \, \e^{+2 V_a} \wt{Q}_a
\Big\}
\nn \\
\ & \ \ \ \ 
+ \int \d^4 \theta \, 
\frac{g^2}{2} \Big\{
\Big( \Gamma + \ol{\Gamma} + 2 \sum_{a=1}^k V_a \Big)^2
- \Big( \Xi + \ol{\Xi} - \sqrt{2} \sum_{a=1}^k (C_a + \ol{C}{}_a) \Big)^2
\Big\}
\nn \\
\ & \ \ \ \ 
+ \sum_{a=1}^k \Big\{
\sqrt{2} \int \d^2 \theta \, \big( \wt{Q}_a \Phi_a Q_a + s_a \, \Phi_a \big)
+ \text{(h.c.)} 
\Big\}
+ \sum_{a=1}^k \Big\{ 
\sqrt{2} \int \d^2 \wt{\theta} \, t_a \Sigma_a
+ \text{(h.c.)}
\Big\}
\nn \\
\ & \ \ \ \ 
- \sqrt{2} \int \d^4 \theta \, 
(\Psi - \ol{\Psi}) \sum_{a=1}^k (C_a - \ol{C}{}_a)
- \sqrt{2} \, \ve^{mn} \sum_{a=1}^k \del_m (\vartheta A_{a,n})
\, . \label{GLSM-522}
\end{align}
This is the model which we proposed in \cite{Kimura:2013fda}.
Here 
$C_a$ is an unconstrained complex superfield related to the chiral superfield $\Phi_a$
in such a way as $\Phi_a = \ol{D}{}_+ \ol{D}{}_- C_a$.
$\Xi$ is an $\N=(2,2)$ twisted chiral superfield dualized from the chiral superfield $\Psi$ under the relation
\begin{align}
\Psi + \ol{\Psi}
- \sqrt{2} \, g^2 \sum_{a=1}^k (C_a + \ol{C}{}_a)
\ &= \ 
- g^2 (\Xi + \ol{\Xi})
\, . \label{Psi2Xi}
\end{align}
The reason why $\Psi - \ol{\Psi}$ still exists in (\ref{GLSM-522}) is because the Lagrangian (\ref{GLSM-KKM}) involves the imaginary part of $\Psi$ as well as its real part. 
Only the real part is explicitly written by the real part of other superfields $\Xi$ and $C_a$ via (\ref{Psi2Xi}) 
under the conventional duality transformation \cite{Rocek:1991ps},
whilst the imaginary part still remains.
Such a phenomenon does not occur in the dualization of (\ref{GLSM-HM}) to (\ref{GLSM-KKM}) 
since the imaginary part of $\Theta$ is not involved in (\ref{GLSM-HM}).
However, we should notice that the existence of $\Psi - \ol{\Psi}$ is not pathological but inevitable 
to realize the background geometry of the exotic $5^2_2$-brane in the IR limit.

We analyze the supersymmetric low energy effective theory of the GLSM (\ref{GLSM-522}) in the IR limit (for the details, see \cite{Kimura:2013fda}).
Under the supersymmetry condition,
the component fields of the charged hypermultiplets $(Q_a, \wt{Q}_a)$ are constrained.
Solving the constraints, we find that they are given in terms of the component fields of the neutral hypermultiplet $(\Xi, \Gamma)$.
Since the gauge coupling constants $e_a$ have mass dimension one,
the vector multiplets $(V_a, \Phi_a)$ become non-dynamical in the IR limit.
If we integrate out all the component fields of the vector multiplets,
we obtain the following low energy effective Lagrangian:
\begin{align}
\Scr{L}_{\text{Eb}}
\ &= \ 
- \frac{1}{2} H \Big\{ (\del_m r^1)^2 + (\del_m r^2)^2 + (\del_m r^3)^2 \Big\}
- \frac{1}{2 H} (\del_m \wt{\vartheta})^2 
\nn \\
\ & \ \ \ \ 
- \frac{({\omega}_{2})^2}{2 H} (\del_m r^2)^2
+ \frac{{\omega}_{2}}{H} (\del_m r^2) (\del^m \wt{\vartheta}) 
+ \ve^{mn} (\del_m r^2) (\del_n y^2) 
\nn \\
\ & \ \ \ \ 
- \frac{({\omega}_{1})^2}{2 H} (\del_m r^1)^2
- \frac{{\omega}_{1} {\omega}_{2}}{H} (\del_m r^1) (\del^m r^2)
+ \frac{{\omega}_{1}}{H} (\del_m r^1) (\del^m \wt{\vartheta}) 
\nn \\
\ & \ \ \ \ 
- \sqrt{2} \, \ve^{mn} \del_m ((\vartheta - t_2) A_{n})
\, , \label{GLSM-522-b1}
\end{align}
where we have omitted the fermionic part.
The explicit forms of the various functionals in (\ref{GLSM-522-b1}) are 
\bsubeq \label{522-fns}
\begin{align}
H \ &= \ 
\frac{1}{g^2} + \sum_{a=1}^k \frac{1}{\sqrt{2} R_a}
\, , \ls
{\omega}_i \ = \ 
\sum_{a=1}^k {\omega}_{i,a} 
\, , \label{sums-fns} \\
R_a \ &= \ 
\sqrt{(r^1 - s_{1,a})^2 + (r^2 - s_{2,a})^2 + (r^3 - t_{1,a})^2}
\, , \\
{\omega}_{1,a} \ &= \ 
\frac{r^2 - s_{2,a}}{\sqrt{2} R_a (R_a + (r^3 - t_{1,a}))}
\, , \ls
{\omega}_{2,a} \ = \ 
- \frac{r^1 - s_{1,a}}{\sqrt{2} R_a (R_a + (r^3 - t_{1,a}))}
\, , \\
t_2 A_m \ &= \ 
\sum_{a=1}^k t_{2,a} A_{m,a}
\ = \ 
\sum_{a=1}^k t_{2,a} \Big\{
\frac{1}{2 R_a H} \Big( \del_m \wt{\vartheta} - {\omega}_i \del_m {r}^i \Big)
+ \frac{1}{\sqrt{2}} {\omega}_{i,a} \del_m {r}^i
\Big\}
\, . 
\end{align}
\esubeq
We should notice that (\ref{GLSM-522-b1}) is not the final
description of the IR effective theory 
as the string worldsheet sigma model for the exotic $5^2_2$-brane of codimension {\it two}.
There are two reasons: 
one is that the target space geometry of (\ref{GLSM-522-b1}) does possess only one isometry.
The other is that the field $r^2$ before the duality transformation
(\ref{Psi2Xi}) still remains caused by the existence of $\Psi - \ol{\Psi}$ in (\ref{GLSM-522}).
In order that we find the genuine effective theory, 
we have to take the following two steps: 
\begin{itemize}
\item[(i)] compactify the $r^2$-direction on $S^1$ with radius ${\cal R}_8$
(i.e., set the FI parameter $s_{2,a}$ to $2 \pi {\cal R}_8 a$), 
and take the infinity limit $k \to \infty$.
\item[(ii)] integrate out the field $r^2$.
\end{itemize}
This is followed from the procedure in the supergravity picture \cite{deBoer:2010ud, deBoer:2012ma}. 
To simplify the configuration, we set $s_{1,a} = 0 = t_{1,a} = t_{2,a}$.
Performing the step (i), we can reduce the sums in (\ref{sums-fns}) to
\bsubeq \label{H-omega-kinfty}
\begin{gather}
H \ = \ 
\frac{1}{g^2} + \sum_{a=1}^k \frac{1}{\sqrt{2} R_a}
\ \xrightarrow{k\to\infty} \ 
H_{\varrho} \ = \ 
\frac{1}{g^2} + \sigma \log \frac{\Lambda}{\varrho}
\ = \ 
h_0 + \sigma \log \frac{\mu}{\varrho}
\, , \\
\omega_1 
\ = \ 
\sum_{a=1}^k \omega_{1,a}
\ \xrightarrow{k\to\infty} \ 0
\, , \ls
\omega_2 
\ = \ 
\sum_{a=1}^k \omega_{2,a}
\ \xrightarrow{k\to\infty} \ 
\omega_{\varrho} 
\ = \ 
\sigma \arctan \Big( \frac{r^3}{r^1} \Big)
\, , \\
\sigma \ = \ \frac{1}{\sqrt{2} \, \pi {\cal R}_8}
\, , \ls
\varrho \ = \ \sqrt{(r^1)^2 + (r^3)^2}
\, .
\end{gather}
\esubeq
Here we introduced the IR cutoff $\Lambda$ because the dimension of the target space is reduced from three to two.
This IR cutoff $\Lambda$ has been regularized by the renormalization scale $\mu$, and $h_0$ is the ``bare'' quantity which diverges in the IR limit.
We refer to this reduction as the smearing procedure \cite{Kimura:2013fda}.
Performing the step (ii), we finally obtain 
\begin{align}
\Scr{L}_{\text{Eb}}
\ &= \ 
- \frac{1}{2} H_{\varrho} \Big\{ (\del_m r^1)^2 + (\del_m r^3)^2 \Big\}
- \frac{H_{\varrho}}{2 K_{\varrho}} 
\Big\{ (\del_m y^2)^2 + (\del_m \wt{\vartheta})^2 \Big\}
\nn \\
\ & \ \ \ \ 
- \frac{\omega_{\varrho}}{K_{\varrho}} \, \ve^{mn} (\del_m y^2) (\del_n \wt{\vartheta})
- \sqrt{2} \, \ve^{mn} \del_m (\vartheta {A}_{n})
\, , \label{NLSM-522}
\end{align}
where $K_{\varrho} = (H_{\varrho})^2 + (\omega_{\varrho})^2$.
This is the NLSM of the background geometry of the exotic $5^2_2$-brane in the presence of the NS-NS B-field \cite{Kimura:2013fda}.
Compared with the string worldsheet sigma model (\ref{string-NLSM}),
we can read off the spacetime variables as follows (for the indices, see Table \ref{ST-WS}):
\begin{gather}
G_{66} \ = \ G_{77} \ = \ 
H_{\varrho}
\, , \ls
G_{88} \ = \ G_{99} \ = \ 
\frac{H_{\varrho}}{K_{\varrho}}
\, , \ls
B_{89} \ = \ 
- \frac{\omega_{\varrho}}{K_{\varrho}}
\, . \label{G-B-522}
\end{gather}

We have demonstrated that the GLSM (\ref{GLSM-522}) becomes the string worldsheet sigma model of the exotic $5^2_2$-brane in the IR limit.
Therefore, as in the similar way as the GLSM for the H-monopoles and for the KK-monopoles \cite{Tong:2002rq, Harvey:2005ab}, 
we can interpret that the GLSM (\ref{GLSM-522}) is a powerful model to analyze the quantum aspects of the string worldsheet sigma model of the exotic $5^2_2$-brane.
One of the concrete analysis is the worldsheet instanton corrections to the background geometry of the exotic $5^2_2$-brane
in the language of the vortex corrections to the GLSM \cite{Kimura:2013zva}.
However, we should notice that 
the GLSM (\ref{GLSM-522}) is sensitive only to the quantum corrections to the topological term involving $\vartheta$.
This implies that, as far as (\ref{GLSM-522}) is concerned, 
we can pursue the quantum corrections only to the $X^9$-direction.
This is, from the viewpoint of the geometry of the exotic $5^2_2$-brane,
the direction along the winding coordinate against the physical coordinate $\wt{X}^9$ (see Table \ref{ST-WS}).

It is noticeable that 
the coordinate fields $y^2$ and $\wt{\vartheta}$ is democratically involved
in the NLSM (\ref{NLSM-522}).
This denotes that the physical coordinates $\wt{X}^8$ and $\wt{X}^9$ in the background geometry of the exotic $5^2_2$-brane.
Then it is natural to discuss quantum corrections to the $\wt{X}^8$-direction or its {\rm T}-dualized direction $X^8$.
In order to trace the quantum corrections to such directions,
we should remodel the GLSM from (\ref{GLSM-522}).
Concretely, we should introduce another vector multiplet coupled to the neutral hypermultiplet $(\Xi, \Gamma)$.
In the next section we will investigate the remodeling of the GLSM (\ref{GLSM-522}).

Before going to the next section,
we have a technical comment on the $k$ sets of $(V_a, \Phi_a; Q_a, \wt{Q}_a)$ in the GLSMs (\ref{GLSM-KKM}) and (\ref{GLSM-522}).
In the spacetime perspective,
the infinity limit $k \to \infty$ gives rise to 
the introduction of an infinite number of five-branes.
The five-branes arrayed in a specific direction generate an isometry \cite{deBoer:2010ud}.
This infinity limit is required to perform the {\rm T}-duality transformation.
In the GLSM perspective, however, this infinity limit seems fearful.
In the IR limit of the GLSM, we solve the equations of motion for the infinite number of $(V_a, \Phi_a; Q_a, \wt{Q}_a)$.
It is no problem in the classical level. 
In the quantum level, however, 
it is unclear whether we can correctly evaluate the path-integral measure of the GLSM with $U(1)^{\infty}$ gauge symmetries.
Then we would like to think of another analysis which involves the corresponding computation under the infinity limit $k \to \infty$.
Now we focus on (\ref{H-omega-kinfty}) discussed above.
This is the procedure of the infinity limit.
More explicitly, we set the FI parameter $s_{2,a}$ to $2 \pi {\cal R}_8 a$ and replace the summations in (\ref{sums-fns}) by the integral under the $k \to \infty$ limit.
Under this procedure we can make another isometry along the $X^8$-direction.
Fortunately, the expression (\ref{sums-fns}) denotes how to realize the same result (\ref{H-omega-kinfty}) even in the $k=1$ system.
We just integrate $1/R$ and $\omega_i$ with respect to the FI parameter $s_2$ in the $k = 1$ system. 
We will mention this procedure more concretely in the next section.

\section{Remodeled GLSM for exotic five-brane}
\label{sect-2-gauged}

There are two isometries on the background geometry of the exotic $5^2_2$-brane.
The GLSM (\ref{GLSM-522}) represents only one of the two isometries being gauged.
It is natural to think of a model in which the other isometry is also gauged.
In this section we consider an extension of the GLSM (\ref{GLSM-522}) by introducing another set of a vector multiplet $(V', \Phi')$ and a charged hypermultiplet $(Q', \wt{Q}')$.

It is easy to add the new multiplets $(V', \Phi'; Q', \wt{Q}')$ to the GLSM (\ref{GLSM-522}).
The new vector superfield $V'$ 
(or $\Sigma' = \frac{1}{\sqrt{2}} \ol{D}{}_+ D_- V'$)
is coupled to the twisted chiral superfield $\Xi$ in the twisted F-term.
The new chiral superfield $\Phi'$ is coupled to the chiral superfield $\Gamma$ in the F-term.
The charged chiral superfields $(Q', \wt{Q}')$ are coupled to $(V', \Phi')$. 
These couplings are standard as in the original GLSM (\ref{GLSM-HM}).
Now we extend the GLSM (\ref{GLSM-522}) in the following form:
\bsubeq \label{X8X9-GLSM}
\begin{align}
\Scr{L}_{\text{E2}}
\ &= \ 
\Scr{L}_{\text{E1}}
+ \Scr{L}_{\text{G}}
\, , \\
\Scr{L}_{\text{E1}}
\ &= \ 
\int \d^4 \theta \, \Big\{ 
\frac{1}{e^2} 
\Big( - \ol{\Sigma} \Sigma + \ol{\Phi} \Phi \Big)
+ \ol{Q} \, \e^{-2 V} Q
+ \ol{\wt{Q}} \, \e^{+2 V} \wt{Q}
\Big\}
\nn \\
\ & \ \ \ \ 
+ \int \d^4 \theta \, 
\frac{g^2}{2} \Big\{ 
\Big( \Gamma + \ol{\Gamma} + 2 V \Big)^2
- \Big( \Xi + \ol{\Xi} - \sqrt{2} (C + \ol{C}) \Big)^2
\Big\}  
\nn \\
\ & \ \ \ \ 
+ \Big\{
\sqrt{2} \int \d^2 \theta \, \big( \wt{Q} \Phi Q + s \, \Phi \big)
+ \text{(h.c.)} 
\Big\}
+ \Big\{ 
\sqrt{2} \int \d^2 \wt{\theta} \, t \, \Sigma
+ \text{(h.c.)}
\Big\}
\nn \\
\ & \ \ \ \ 
- \sqrt{2} \int \d^4 \theta \, 
(\Psi - \ol{\Psi}) (C - \ol{C})
- \sqrt{2} \, \ve^{mn} \del_m (\vartheta A_{n})
\, , \label{X8X9-522} \\
\Scr{L}_{\text{G}}
\ &= \ 
\int \d^4 \theta \, \Big\{
\frac{1}{e'{}^2} \Big( - \ol{\Sigma}{}' \Sigma' + \ol{\Phi}{}' \Phi' \Big)
+ \ol{Q}{}' \, \e^{-2 V'} Q'
+ \ol{\wt{Q}}{}' \, \e^{+2 V'} \wt{Q}'
\Big\}
\nn \\
\ & \ \ \ \ 
+ \Big\{
\sqrt{2} \int \d^2 \theta \, \big( \wt{Q}' \Phi' Q' 
+ (s' - \Gamma) \Phi' \big)
+ \text{(h.c.)} 
\Big\}
\nn \\
\ & \ \ \ \ 
+ \Big\{ 
\sqrt{2} \int \d^2 \wt{\theta} \, (t' - \Xi) \Sigma'
+ \text{(h.c.)}
\Big\}
\, . \label{X8X9-add}
\end{align}
\esubeq
Here $\Scr{L}_{\text{E1}}$ is the GLSM (\ref{GLSM-522}) with $k=1$,
whilst $\Scr{L}_{\text{G}}$ is the additional part to the GLSM given by the new multiplets $(V', \Phi'; Q', \wt{Q}')$.
All the gauge couplings in $\Scr{L}_{\text{G}}$ are minimal as we can see in the original GLSM (\ref{GLSM-HM}).
We remark that $e'$ is the gauge coupling constant of the additional vector multiplet $(V', \Phi')$, 
and $(s', t')$ are the additional complex-valued FI parameters. 
In the present paper, 
we refer to the model (\ref{X8X9-GLSM}) as the Remodeled GLSM for the exotic $5^2_2$-brane.
The duality transformed models of (\ref{X8X9-GLSM}) are briefly discussed in appendix \ref{app-KKM2-H2}.

We have a comment on supersymmetry of the Remodeled GLSM (\ref{X8X9-GLSM}).
The Remodeled GLSM (\ref{X8X9-GLSM}) has $\N=(2,2)$ supersymmetry rather than $\N = (4,4)$ supersymmetry 
caused by a difficulty of the assignment of $SU(2)_R$ symmetry:
In the Remodeled GLSM (\ref{X8X9-GLSM}), 
the neutral hypermultiplet $(\Xi,\Gamma)$ is coupled to not only the original vector multiplet $(V, \Phi)$ but also the additional one $(V', \Phi')$. 
The former multiplet assigns the three scalar fields $(r^1, r^2, r^3)$ in the neutral hypermultiplet to the triplet of an $SU(2)_R$ \cite{Tong:2002rq, Harvey:2005ab},
whilst the latter multiplet also assigns the three scalar fields $(r^1, r^3, \gamma^4)$ to the triplet of another $SU(2)_R$ (for details, see later discussions).
Since it is hard to preserve both the $SU(2)_R$ symmetries simultaneously,
$\N=(4,4)$ supersymmetry in $\Scr{L}_{\text{E1}}$ 
is broken down to $\N=(2,2)$ supersymmetry when $\Scr{L}_{\text{G}}$ is added.
Of course the $\N=(2,2)$ supersymmetry is manifest because the model
(\ref{X8X9-GLSM}) is formulated in terms of $\N=(2,2)$ superfields.
If we adopt a certain condition in the supersymmetric vacua,
the $\N=(4,4)$ supersymmetry would be restored.

In the rest of this section 
we focus on the bosonic sector and investigate the classical structure of the Remodeled GLSM (\ref{X8X9-GLSM}).
In the next section we will study the quantum aspects of the Remodeled GLSM.

\subsection{Bosonic Lagrangian}

In this subsection we carefully compute the bosonic sector of the Remodeled GLSM (\ref{X8X9-GLSM}) 
because this is more complicated than the GLSM (\ref{GLSM-522}).
First, we expand all the superfields in the Remodeled GLSM (\ref{X8X9-GLSM})
in the presence of auxiliary fields.
Second, we integrate out all the auxiliary fields and obtain the bosonic Lagrangian with constraints.

Following the expansion rule in appendix \ref{app-SF},
we write down the bosonic sector of (\ref{X8X9-GLSM}):
\begin{align}
\Scr{L}_{\text{E2b}}
\ &= \ 
\frac{1}{e^2} \Big\{
\half (F_{01})^2 
- |\del_m \sigma|^2
- 4 |\del_m M_{c}|^2
\Big\}
+ \frac{1}{e'{}^2} \Big\{ \half (F'_{01})^2  
- |\del_m \sigma'|^2 - |\del_m \phi'|^2 \Big\}
\nn \\
\ & \ \ \ \ 
- \frac{1}{2 g^2} \Big\{ (\del_m r^1)^2 + (\del_m r^3)^2 \Big\}
- \frac{g^2}{2} \Big\{ (\del_m y^2)^2 + (D_m \gamma^4)^2 \Big\}
\nn \\
\ & \ \ \ \ 
- \Big\{ |D_m q|^2 + |D_m \wt{q}|^2 \Big\}
- \Big\{ |D_m q'|^2 + |D_m \wt{q}'|^2 \Big\}
- \sqrt{2} \, \ve^{mn} \del_m \big( (\vartheta - t_{2}) A_{n} \big)
- \sqrt{2} (y^2 - t'_{2}) \, F'_{01}
\nn \\
\ & \ \ \ \
- 2 |\sigma|^2 \big( |q|^2 + |\wt{q}|^2 + g^2 \big)
-2 |\sigma'|^2 \big( |q'|^2 + |\wt{q}'|^2 \big)
\nn \\
\ & \ \ \ \ 
+ |F|^2 + |\wt{F}|^2
+ 2 \sqrt{2} M_{c} \big( q \wt{F} + \wt{q} F \big)
+ 2 \sqrt{2} \ol{M}{}_{c} \big( \ol{q} \ol{\wt{F}} + \ol{\wt{q}} \ol{F} \big)
\nn \\
\ & \ \ \ \ 
- 2 g^2  |F_{c} + \ol{M}{}_{c}|^2
+ 4 g^2 \big( |F_{c}|^2 - |M_{c}|^2 \big)
+ {g^2} |{G}_{\Gamma}|^2
- \i \sqrt{2} \Big\{
\big( \phi' \, G_{\Gamma} - \ol{\phi}{}' \, \ol{G}{}_{\Gamma} \big)
+ \big( \sigma' G_{\Xi} - \ol{\sigma}{}' \ol{G}{}_{\Xi} \big)
\Big\}
\nn \\
\ & \ \ \ \ 
+ |F'|^2 + |\wt{F}'|^2
+ \i \sqrt{2} \, \phi' \big( q' \wt{F}' + \wt{q}' F' \big)
- \i \sqrt{2} \, \ol{\phi}{}' \big( \ol{q}{}' \ol{\wt{F}}{}' + \ol{\wt{q}}{}' \ol{F}{}' \big)
\nn \\
\ & \ \ \ \ 
+ \frac{1}{2 e^2} (D_{V})^2
+ D_{V} \big( |q|^2 - |\wt{q}|^2 - \sqrt{2} \, (r^3 - t_{1}) \big)
\nn \\
\ & \ \ \ \
+ \frac{1}{2 e'{}^2} (D'_{V})^2 
+ D'_{V} \Big\{
|q'|^2 - |\wt{q}'|^2 
+ \sqrt{2} \Big( \frac{r^1}{g^2} + t'_{1} - ( \phi_{c} + \ol{\phi}{}_{c} ) \Big)
\Big\}
\nn \\
\ & \ \ \ \ 
+ \frac{1}{e'{}^2} |D'_{\Phi}|^2
+ \i D'_{\Phi} \Big\{
\sqrt{2} \, q' \wt{q}' 
- \Big( - \frac{r^3}{g^2} + \i \gamma^4 - \sqrt{2} \, s' \Big)
\Big\}
- \i \ol{D}{}'_{\Phi} \Big\{
\sqrt{2} \, \ol{q}{}' \ol{\wt{q}}{}'
- \Big( - \frac{r^3}{g^2} - \i \gamma^4 - \sqrt{2} \, \ol{s}{}' \Big)
\Big\}
\nn \\
\ & \ \ \ \ 
+ \frac{1}{e^2} 
\big| D_{c} + \sqrt{2} \, e^2 \, \ol{q} \ol{\wt{q}} \big|^2
- 2 e^2 |q \wt{q}|^2
- D_{c} \big( (r^1 - s_{1}) + \i (r^2 - s_{2}) \big)
- \ol{D}{}_{c} \big( (r^1 - s_{1}) - \i (r^2 - s_{2}) \big)
\nn \\
\ & \ \ \ \ 
- \frac{\i}{2 e^2} 
\big( D_{c} + \sqrt{2} \, e^2 \, \ol{q} \ol{\wt{q}} \big)
\Big\{ (\del_0 - \del_1) \ol{B}{}_{c\+}
+ \i (\del_0 + \del_1) \ol{A}{}_{c=}
- \i (\del_0^2 - \del_1^2) \ol{\phi}{}_{c} \Big\}
\nn \\
\ & \ \ \ \ 
+ \frac{\i}{2 e^2} 
\big( \ol{D}{}_{c} + \sqrt{2} \, e^2 \, q \wt{q} \big)
\Big\{ (\del_0 - \del_1) B_{c\+}
- \i (\del_0 + \del_1) A_{c=}
+ \i (\del_0^2 - \del_1^2) \phi_{c} \Big\}
\nn \\ 
\ & \ \ \ \  
- \frac{g^2}{2} \Big\{
(B_{c\+} + \ol{B}{}_{c\+}) (\del_0 - \del_1) y^2
- (A_{c=} + \ol{A}{}_{c=}) (\del_0 + \del_1) y^2
\Big\}
\nn \\
\ & \ \ \ \ 
+ \frac{\i}{2} (\phi_{c} - \ol{\phi}{}_{c}) (\del_0^2 - \del_1^2) r^2
- \frac{g^2}{2} 
(A_{c=} + \ol{A}{}_{c=}) (B_{c\+} + \ol{B}{}_{c\+})
\nn \\
\ & \ \ \ \ 
+ \frac{\i}{2} \Big\{
(B_{c\+} - \ol{B}{}_{c\+}) (\del_0 - \del_1) r^1
- (A_{c=} - \ol{A}{}_{c=}) (\del_0 + \del_1) r^1 \Big\}
\nn \\
\ & \ \ \ \ 
+ \half (\phi_{c} + \ol{\phi}{}_{c}) (\del_0^2 - \del_1^2) r^1 
+ \frac{1}{4 e^2} \Big|
(\del_0 - \del_1) B_{c\+}
- \i (\del_0 + \del_1) A_{c=}
+ \i (\del_0^2 - \del_1^2) \phi_{c}
\Big|^2
\, . \label{X8X9-GLSM-b1}
\end{align}
Here the gauge covariant derivatives in the above Lagrangian are
\bsubeq
\begin{alignat}{3}
D_m q \ &= \ 
\del_m q - \i A_{m} q
\, , &\ls
D_m \wt{q}
\ &= \ 
\del_m \wt{q} + \i A_{m} \wt{q}
\, , &\ls
D_m \gamma^4 \ &= \ 
\del_m \gamma^4 + \sqrt{2} A_{m}
\, , \\
D_m q' \ &= \ 
\del_m q' - \i A'_{m} \, q'
\, , &\ls
D_m \wt{q}'  \ &= \ 
\del_m \wt{q}' + \i A'_{m} \, \wt{q}'
\, .  
\end{alignat}
\esubeq
The covariant derivatives in the second line are originated from $\Scr{L}_{\text{G}}$ (\ref{X8X9-add}).
The Lagrangian (\ref{X8X9-GLSM-b1}) contains the terms involving the scalar field $r^2$ (rather than the derivative $\del_m r^2$). 
They are originated from $(\Psi - \ol{\Psi})(C - \ol{C})$ in $\Scr{L}_{\text{E1}}$ (\ref{X8X9-522}). 
The scalar field $r^2$ is the dual field of $y^2$ via the following forms by the duality relation (\ref{Psi2Xi}) \cite{Kimura:2013fda}:
\bsubeq \label{const-DualSFs}
\begin{align}
(\del_0 + \del_1) r^2 \ &= \ 
- g^2 (\del_0 + \del_1) y^2 
+ g^2 \big( B_{c\+} + \ol{B}{}_{c\+} \big)
\, , \\
(\del_0 - \del_1) r^2 \ &= \ 
+ g^2 (\del_0 - \del_1) y^2 
+ g^2 \big( A_{c=} + \ol{A}{}_{c=} \big)
\, .
\end{align}
\esubeq
The Lagrangian (\ref{X8X9-GLSM-b1}) involves the following 
auxiliary fields of the superfield formalism: 
\begin{align*}
D_{V} \, , \ D_{c} \, , \ 
D'_{V} \, , \ D'_{\Phi} \, , \ 
G_{\Gamma} \, , \ G_{\Xi} \, , \ 
F \, , \ \wt{F} \, , \ 
F' \, , \ \wt{F}' \, , \ 
F_{c} \, , \ A_{c=} \, , \ B_{c\+} \, , \ \phi_{c} 
\, .
\end{align*}
Integrating them out, we obtain the bosonic part of the Lagrangian represented only in terms of dynamical fields:
\bsubeq \label{X8X9-GLSM-b2}
\begin{align}
\Scr{L}_{\text{E2b}}
\ &= \ 
\frac{1}{e^2} \Big\{
\half (F_{01})^2 
- |\del_m \sigma|^2
- 4 |\del_m M_{c}|^2
\Big\}
+ \frac{1}{e'{}^2} \Big\{
\half (F'_{01})^2 
- |\del_m \sigma'|^2
- |\del_m \phi'|^2
\Big\}
\nn \\
\ & \ \ \ \ 
- \frac{1}{2 g^2} \Big\{ (\del_m r^1)^2 + (\del_m r^3)^2 \Big\}
- \frac{g^2}{2} \Big\{ (\del_m y^2)^2 + (D_m \gamma^4)^2 \Big\}
\nn \\
\ & \ \ \ \ 
- \Big\{
|D_m q|^2
+ |D_m \wt{q}|^2
\Big\}
- \Big\{ |D_m q'|^2 + |D_m \wt{q}'|^2 \Big\}
- \sqrt{2} \, \ve^{mn} \del_m \big( (\vartheta - t_{2}) A_{n} \big)
- \sqrt{2} (y^2 - t'_{2}) F'_{01}
\nn \\
\ & \ \ \ \ 
- 2 \big( |\sigma|^2 + 4 |M_{c}|^2 \big) 
\big( |q|^2 + |\wt{q}|^2 + g^2 \big)
- 2 |\phi'|^2 \Big( |q'|^2 + |\wt{q}'|^2 + \frac{1}{g^2} \Big)
\nn \\
\ & \ \ \ \ 
- \frac{e^2}{2} \big( |q|^2 - |\wt{q}|^2 - \sqrt{2} \, (r^3 - t_{1}) \big)^2
- e^2 \big| \sqrt{2} \, q \wt{q} - \big( (r^1 - s_{1}) + \i (r^2 - s_{2}) \big) \big|^2
\nn \\
\ & \ \ \ \ 
- e'{}^2 \Big| \sqrt{2} \, q' \wt{q}' 
- \Big( - \frac{r^3}{g^2} + \i \gamma^4 - \sqrt{2} \, s' \Big) \Big|^2
\nn \\
\ & \ \ \ \ 
+ \frac{g^2}{2} 
(A_{c=} + \ol{A}{}_{c=}) (B_{c\+} + \ol{B}{}_{c\+})
\, . \label{X8X9-GLSM-b21}
\end{align}
Notice that the Lagrangian is constrained by the following equation 
caused by the equation of motion for the auxiliary field $D'_V$:
\begin{align}
0 \ &= \ 
|q'|^2 - |\wt{q}'|^2 
+ \sqrt{2} \Big( \frac{r^1}{g^2} + t'_{1} 
- (\phi_{c} + \ol{\phi}{}_{c}) \Big)
\, . \label{const-add-X8X9-GLSM}
\end{align}
\esubeq
Due to the duality relation (\ref{Psi2Xi})
and the equations of motion for the auxiliary fields $A_{c=}$ and $B_{c\+}$,
the last line in the right-hand side of (\ref{X8X9-GLSM-b21}) is expressed in the following way:
\begin{align}
\frac{g^2}{2} 
(A_{c=} &+ \ol{A}{}_{c=}) (B_{c\+} + \ol{B}{}_{c\+})
\nn \\
\ &= \ 
- \frac{1}{2 g^2} (\del_m r^2)^2
+ \frac{g^2}{2} (\del_m y^2)^2
+ \ve^{mn} (\del_m r^2) (\del_n y^2)
\, . \label{const-add2-X8X9-GLSM}
\end{align}
Plugging this into (\ref{X8X9-GLSM-b2}), 
we find that the kinetic term of $y^2$ disappears and the kinetic term of the dual field $r^2$ is revived.
Even though this phenomenon looks strange,
we should keep in mind that the dynamical field is $y^2$ rather than $r^2$ in the system. 
We should integrate out $r^2$ in the final stage of the analysis \cite{Kimura:2013fda}.

\subsection{Low energy limit}

In this subsection we investigate the supersymmetric low energy effective theory in the IR limit.
We are interested in the Higgs branch of the model where scalar fields
of the charged hypermultiplets have non-trivial vacuum expectation values as discussed in \cite{Kimura:2013fda}.
We first evaluate the supersymmetric vacua from the vanishing condition of the potential terms.
Second, taking the IR limit $e, e' \to \infty$,
we remove the kinetic terms of the gauge fields $A_m$ and $A'_m$.
Third, we perform the smearing procedure which generates two isometries of the target space of the sigma model. 
Finally, integrating out the gauge fields and the dual field $r^2$, 
we obtain the NLSM which has only $\N=(2,2)$ supersymmetry.

\subsubsection*{Solving constraints on charged hypermultiplets}

We evaluate the supersymmetric vacua obtained by the vanishing condition of the potential terms of (\ref{X8X9-GLSM-b2}).
Since we are interested in the Higgs branch, 
the vanishing condition is given by the following configuration:
\bsubeq \label{SUSYvacua-X8X9-GLSM}
\begin{alignat}{2}
0 \ &= \ 
\sigma \ = \ M_{c} 
\, , &\ls
0 \ &= \ 
\phi'
\, , \ls
0 \ = \ 
\phi_{c}
\, , \label{noCoulomb} \\
0 \ &= \ 
|q|^2 - |\wt{q}|^2
- \sqrt{2} (r^3 - t_{1})
\, , &\ls 
0 \ &= \ 
\sqrt{2} \, q \wt{q}
- \big( (r^1 - s_{1}) + \i (r^2 - s_{2}) \big)
\, , \label{constqq} \\
0 \ &= \ 
|q'|^2 - |\wt{q}'|^2 
- \sqrt{2} \Big( - \frac{r^1}{g^2} - t'_{1} \Big)
\, , &\ls
0 \ &= \ 
\sqrt{2} \, q' \wt{q}'
- \Big( - \frac{r^3}{g^2} + \i \gamma^4 - (s'_{1} + \i s'_{2}) \Big)
\, . \label{constq'q'}
\end{alignat}
\esubeq
The first line (\ref{noCoulomb}) implies that all the scalar fields in the vector multiplet are trivial.
The equations (\ref{constqq}) constrain the scalar fields of the charged hypermultiplet $(Q, \wt{Q})$ by the scalar fields of the neutral hypermultiplet $(\Xi, \Gamma)$.
The solution is 
\bsubeq \label{qqt-X8X9-GLSM}
\begin{gather}
q \ = \ 
- \frac{\i}{2^{1/4}} \, \e^{- \i \alpha}
\sqrt{R + (r^3 - t_{1})}
\, , \ls
\wt{q} \ = \ 
\frac{\i}{2^{1/4}} \, \e^{+ \i \alpha}
\frac{(r^1 - s_{1}) + \i (r^2 - s_{2})}{\sqrt{R + (r^3 - t_{1})}}
\, , \\
R \ = \ 
\sqrt{(r^1 - s_{1})^2 + (r^2 - s_{2})^2 + (r^3 - t_{1})^2}
\, . 
\end{gather}
\esubeq
Plugging this into the kinetic terms of the scalar fields $(q, \wt{q})$, we obtain
\bsubeq \label{DqDqt-X8X9-GLSM}
\begin{gather}
\begin{align}
- |D_m q|^2 - |D_m \wt{q}|^2
\ &= \ 
- \frac{1}{2 \sqrt{2} R}
\Big\{ (\del_m r^1)^2 + (\del_m r^2)^2 + (\del_m r^3)^2 \Big\}
\nn \\
\ & \ \ \ \ 
- \sqrt{2} R \Big( \del_m \alpha + A_{m} - \frac{1}{\sqrt{2}} \omega_{i} \, \del_m r^i \Big)^2
\, , \\
\omega_{i} \, \del_m r^i
\ &= \ 
\omega_{1} \, \del_m r^1
+ \omega_{2} \, \del_m r^2
+ \omega_{3} \, \del_m r^3
\, , 
\end{align}
\\
\omega_{1}
\ = \ 
\frac{r^2 - s_{2}}{\sqrt{2} R (R + (r^3 - t_{1}))}
\, , \ls
\omega_{2}
\ = \ 
\frac{- (r^1 - s_{1})}{\sqrt{2} R (R + (r^3 - t_{1}))}
\, , \ls
\omega_{3}
\ = \ 
0
\, .
\end{gather}
\esubeq
This form implies that there is a rotational symmetry among $(r^1, r^2, r^3)$ and among $(\omega_1, \omega_2, \omega_3)$. 
This rotational symmetry is an $SU(2)_R$ symmetry of the $\N=(4,4)$ supersymmetry assigned by the vector multiplet $(V, \Phi)$.
This is one of the concrete assignment of the representation of the $SU(2)_R$ symmetry, 
though it is of course possible to impose the same charge assignment of this $SU(2)_R$ symmetry in the GLSM level.

We also solve the equations (\ref{constq'q'}) as follows:
\bsubeq \label{q'q't-X8X9-GLSM}
\begin{gather}
q' \ = \ 
- \frac{\i}{2^{1/4} g} \, \e^{- \i \alpha'}
\sqrt{R' + (- r^1 - g^2 t'_{1})}
\, , \ls
\wt{q}' \ = \ 
\frac{\i}{2^{1/4} g} \, \e^{+ \i \alpha'}
\frac{(- r^3 - g^2 s'_{1}) + \i (g^2 \gamma^4 - g^2 s'_{2})}{\sqrt{R' + (- r^1 - g^2 t'_{1})}}
\, , \\
R' \ = \ 
\sqrt{(- r^1 - g^2 t'_{1})^2
+ (- r^3 - g^2 s'_{1})^2
+ (g^2 \gamma^4 - g^2 s'_{2})^2}
\, .
\end{gather}
\esubeq
Substituting this into the kinetic terms of the scalar fields  $(q', \wt{q}')$, we obtain
\bsubeq \label{Dq'Dq't-X8X9-GLSM}
\begin{gather}
\begin{align}
- |D_m q'|^2 - |D_m \wt{q}'|^2
\ &= \ 
- \frac{1}{2 \sqrt{2} \, g^2 R'}
\Big\{ (\del_m r^1)^2 + (\del_m r^3)^2 + g^4 (\del_m \gamma^4)^2 \Big\}
\nn \\
\ & \ \ \ \
- \frac{\sqrt{2} R'}{g^2} 
\Big( \del_m \alpha' + A'_{m} - \frac{1}{\sqrt{2}} \omega'_{j} \, \del_m \wh{r}^j \Big)^2 
\, , \\
\omega'_{j} \, \del_m \wh{r}^j
\ &= \ 
\omega'_{1} \, \del_m r^1
+ \omega'_{3} \, \del_m r^3
+ \omega'_{4} \, \del_m (g^2 \gamma^4)
\, , 
\end{align}
\\
\omega'_{1} 
\ = \ 0
\, , \ls
\omega'_{3} 
\ = \ 
\frac{- (g^2 \gamma^4 - g^2 s'_{2})}{\sqrt{2} R' (R' + (- r^1 - g^2 t'_{1}))}
\, , \ls
\omega'_{4} 
\ = \ 
\frac{- (- r^3 - g^2 s'_{1})}{\sqrt{2} R' (R' + (- r^1 - g^2 t'_{1}))}
\, .
\end{gather}
\esubeq
The above form also implies that there is a rotational symmetry among $(r^1, r^3, g^2 \gamma^4)$ and among $(\omega'_1, \omega'_3, \omega'_4)$. 
This rotational symmetry is another $SU(2)_R$ symmetry of the $\N=(4,4)$ supersymmetry assigned by the vector multiplet $(V', \Phi')$.
This is one of the concrete assignment of the representation of the second $SU(2)_R$ symmetry.
As mentioned before, 
it is hard to preserve both of the $SU(2)_R$ symmetries simultaneously.
The two $SU(2)_R$ symmetries are broken down to $U(1)_R$ 
and the system (\ref{X8X9-GLSM}) has only $\N=(2,2)$ supersymmetry.

\subsubsection*{Integrating-out of gauge fields}

Now we take the IR limit $e, e' \to \infty$.
In this limit the kinetic terms of the gauge fields $A_m$ and $A'_m$ shrink to zero and they become auxiliary fields.
In order to integrate them out explicitly, 
we first substitute the supersymmetry condition (\ref{SUSYvacua-X8X9-GLSM}), 
the solutions (\ref{qqt-X8X9-GLSM}), (\ref{q'q't-X8X9-GLSM})
and the covariant derivatives (\ref{DqDqt-X8X9-GLSM}), (\ref{Dq'Dq't-X8X9-GLSM})
into the Lagrangian (\ref{X8X9-GLSM-b2}):
\begin{align}
\Scr{L}_{\text{E2b}}
\ &= \ 
- \frac{1}{2 g^2} \Big\{ (\del_m r^1)^2 + (\del_m r^2)^2 + (\del_m r^3)^2 \Big\}
- \frac{g^2}{2} (D_m \gamma^4)^2 
+ \ve^{mn} (\del_m r^2) (\del_n y^2)
\nn \\
\ & \ \ \ \ 
- \frac{1}{2 \sqrt{2} R}
\Big\{ (\del_m r^1)^2 + (\del_m r^2)^2 + (\del_m r^3)^2 \Big\}
- \sqrt{2} R \Big( \del_m \alpha + A_{m} - \frac{1}{\sqrt{2}} \omega_{i} \, \del_m r^i \Big)^2
\nn \\
\ & \ \ \ \ 
- \frac{1}{2 \sqrt{2} \, g^2 R'}
\Big\{ (\del_m r^1)^2 + (\del_m r^3)^2 + g^4 (\del_m \gamma^4)^2 \Big\}
- \frac{\sqrt{2} R'}{g^2} 
\Big( \del_m \alpha' + A'_{m} - \frac{1}{\sqrt{2}} \omega'_{j} \, \del_m \wh{r}^j \Big)^2 
\nn \\
\ & \ \ \ \ 
- \sqrt{2} \, \ve^{mn} \del_m \big( (\vartheta - t_{2}) A_{n} \big)
+ \sqrt{2} \, \ve^{mn} (y^2 - t'_{2}) \, \del_n A'_{m}
\, . \label{X8X9-GLSM-b3}
\end{align}
Then we can solve the equations of motion for the gauge fields $A_{m}$, $A'_m$. 
The solution is
\bsubeq \label{sol-AA'-X8X9}
\begin{align}
A_{m} \ &= \ 
- \frac{1}{\sqrt{2} \, g^2 H} \Big( \del_m \wt{\vartheta} - \omega_i \, \del_m r^i \Big)
+ \frac{1}{\sqrt{2}} \del_m \wt{\vartheta}
- \del_m \alpha 
\, , \label{solA-X8X9-GLSM} \\
A'_{m}
\ &= \ 
\frac{1}{\sqrt{2}} \omega'_{j} \, \del_m \wh{r}^j
+ \frac{g^2}{\sqrt{2}} H' \ve_{mn} \, (\del^n y^2)
- \del_m \alpha' 
\, , \label{solA'-X8X9-GLSM} \\
H \ &= \ 
\frac{1}{g^2} + \frac{1}{\sqrt{2} R}
\, , \ls
H' \ = \ 
\frac{1}{\sqrt{2} R'}
\, .
\end{align}
\esubeq
Here $\alpha$ and $\alpha'$ are the unfixed phase factors of $(q, \wt{q})$ and $(q', \wt{q}')$, respectively.
They can be interpreted as the gauge parameters of $A_m$ and $A'_m$.
We also introduced the representation $\gamma^4 = \wt{\vartheta}$ up to the gauge transformation \cite{Tong:2002rq}.
Imposing the gauge-fixing condition $\alpha = 0 = \alpha'$,
we substitute (\ref{sol-AA'-X8X9}) into (\ref{X8X9-GLSM-b3}):
\begin{align}
\Scr{L}_{\text{E2b}}
\ &= \ 
- \frac{1}{2} \Big( H + \frac{1}{g^2} H' \Big)
\Big\{ (\del_m r^1)^2 + (\del_m r^3)^2 \Big\}
- \frac{1}{2} \Big( \frac{1}{H} + g^2 H' \Big) (\del_m \wt{\vartheta})^2 
\nn \\
\ & \ \ \ \ 
- \frac{\omega_i \omega_j}{2 H} (\del_m r^i) (\del^m r^j)
+ \frac{\omega_i}{H} (\del_m r^i) (\del^m \wt{\vartheta}) 
\nn \\
\ & \ \ \ \ 
- \frac{1}{2} H (\del_m r^2)^2 
- \frac{g^2}{2} H' (\del_m y^2)^2
+ \ve^{mn} \Big( \del_m r^2 + \omega'_j \, \del_m \wh{r}^j \Big) (\del_n y^2) 
\nn \\
\ & \ \ \ \ 
- \sqrt{2} \, \ve^{mn} \del_m \big( (\vartheta - t_{2}) A_{n} \big)
\, . \label{X8X9-GLSM-b4}
\end{align}
This is {\it not} the sigma model for the exotic $5^2_2$-brane
because the dual field $r^2$ still contributes to the system.
In order to generate the genuine background geometry of the exotic $5^2_2$-brane, 
we take the following reduction discussed below.
This is identical to the smearing procedure performed in (\ref{H-omega-kinfty}).

\subsubsection*{Smearing procedure: generating isometries}

We perform the smearing procedure 
identical to the computations (\ref{H-omega-kinfty}).
Integration of the functionals with respect to the FI parameters $s_2$ and $s'_2$ yields the shift symmetries $r^2 \to r^2 + \beta$ and $\wt{\vartheta} \to \wt{\vartheta} + \wt{\beta}$, 
where $\beta$ and $\wt{\beta}$ are arbitrary constants.
First we focus on the functionals $H$ and $\omega_i$.
They contain the field $r^2$ with the FI parameter $s_2$, 
whilst they do not depend on the field $\gamma^4 = \wt{\vartheta}$ and the FI parameter $s'_2$.
The other functionals $H'$ and $\omega'_i$ do not depend on $r^2$ and $s_2$,
whilst they are subject to the field $\gamma^4 = \wt{\vartheta}$ and the FI parameter $s'_2$.

Now we set the FI parameter $s_2$ to $2 \pi {\cal R}_8 s$, 
where ${\cal R}_8$ corresponds to the radius of the compactified direction $r^2 = X^8$.
This is dual to the physical coordinate $\wt{X}^8$ in the background geometry of the exotic $5^2_2$-brane.
We integrate the functionals over the parameter $s$:
\bsubeq \label{8-isometry}
\begin{alignat}{2}
s_{2} \ &= \ 2 \pi {\cal R}_8 \, s
\, , \\
H \ &\xrightarrow{\text{integral over $s$}} \ 
H_{\varrho} \ = \ \frac{1}{g^2} + \sigma \log \frac{\Lambda}{\varrho}
\, , &\ls
\sigma \ &= \ \frac{1}{\sqrt{2} \pi {\cal R}_8}
\, , \\
\omega_1 \ &\xrightarrow{\text{integral over $s$}} \ 0
\, , &\ls
\varrho \ &= \ 
\sqrt{(r^1 - s_1)^2 + (r^3 - t_1)^2}
\, , \\
\omega_2 \ &\xrightarrow{\text{integral over $s$}} \ 
\omega_{\varrho} \ = \ 
\sigma \, \vartheta_{\varrho}
\, , &\ls
\vartheta_{\varrho} 
\ &= \ 
\arctan \Big( \frac{r^3 - t_1}{r^1 - s_1} \Big)
\, .
\end{alignat}
\esubeq
Due to this, all the functionals do not depend on the dual field $r^2$ any more. 
This denotes that the system has an isometry along the $X^8$-direction.
We note that $\Lambda$ is the IR cutoff on the two-dimensional $(r^1, r^3) = (X^6,X^7)$ plane.
In the same way, we set the FI parameter $s'_2$ to $2 \pi \wt{\cal R}_9 s'$, 
where $\wt{\cal R}_9$ corresponds to the compact radius of the physical coordinate $\wt{X}^9$.
We integrate the functionals over the parameter $s'$:
\bsubeq \label{9-isometry}
\begin{alignat}{2}
s'_{2} \ &= \ 2 \pi \wt{\cal R}_9 \, s'
\, , \\
H' \ &\xrightarrow{\text{integral over $s'$}} \ 
H'_{\varrho}
\ = \ 
\sigma' \log \frac{\Lambda'}{\varrho'}
\, , &\ls
\sigma' \ &= \ \frac{1}{\sqrt{2} \pi \wt{\cal R}_9}
\, , \\
\omega'_3 \ &\xrightarrow{\text{integral over $s'$}} \ 0
\, , &\ls
\varrho'
\ &= \ 
\sqrt{(- r^1 - g^2 t'_1)^2 + (- r^3 - g^2 s'_1)^2}
\, , \\
\omega'_4 \ &\xrightarrow{\text{integral over $s'$}} \ 
\frac{1}{g^2} \omega'_{\varrho} 
\ = \ 
\frac{\sigma'}{g^2} \, \vartheta'_{\varrho}
\, , &\ls
\vartheta'_{\varrho}
\ &= \ 
\arctan \Big( \frac{- r^1 - g^2 t'_1}{- r^3 - g^2 s'_1} \Big)
\, .
\end{alignat}
\esubeq
Due to this, the system does not depend on the field $\gamma^4 = \wt{\vartheta}$ any more.
This indicates that the target space geometry of the sigma model has an isometry along the $\wt{X}^9$-direction.
Here we also introduced another IR cutoff parameter $\Lambda'$ of the two-dimensional $(r^1, r^3) = (X^6,X^7)$ plane.
Applying the reductions (\ref{8-isometry}) and (\ref{9-isometry}) to the Lagrangian (\ref{X8X9-GLSM-b4}), we obtain 
\begin{align}
\Scr{L}_{\text{E2b}}
\ &= \ 
- \frac{1}{2} \Big( H_{\varrho} + \frac{1}{g^2} H' \Big)
\Big\{ (\del_m r^1)^2 + (\del_m r^3)^2 \Big\}
- \half \Big( \frac{1}{H_{\varrho}} + g^2 H_{\varrho}' \Big)
(\del_m \wt{\vartheta})^2 
\nn \\
\ & \ \ \ \ 
- \frac{K_{\varrho}}{2 H_{\varrho}} (\del_m r^2)^2
+ \frac{\omega_{\varrho}}{H_{\varrho}} (\del_m r^2) (\del^m \wt{\vartheta}) 
- \frac{g^2}{2} H'_{\varrho} (\del_m y^2)^2
+ \ve^{mn} \Big( \del_m r^2 + \frac{1}{g^2} \omega'_{\varrho} \, \del_m (g^2 \wt{\vartheta}) \Big) (\del_n y^2)
\nn \\
\ & \ \ \ \ 
- \sqrt{2} \, \ve^{mn} \del_m \big( (\vartheta - t_{2}) A_{n} \big)
\, . \label{X8X9-GLSM-b5}
\end{align}
Now we are ready to integrate out the dual field $r^2$ to complete the {\rm T}-duality transformation \cite{Kimura:2013fda}.
The solution of the integration is given by
\begin{align}
\del_m r^2 \ &= \ 
\frac{H_{\varrho}}{K_{\varrho}} \Big\{
\frac{\omega_{\varrho}}{H_{\varrho}} (\del_m \wt{\vartheta}) 
+ \ve_{mn} (\del^n y^2) \Big\}
\, . \label{solr2-X8X9-GLSM}
\end{align}
Plugging this into the above Lagrangian, we obtain the final form of the NLSM:
\begin{align}
\Scr{L}_{\text{E2b}}
\ &= \ 
- \frac{1}{2} \Big( H_{\varrho} + \frac{1}{g^2} H' \Big) 
\Big\{ (\del_m r^1)^2 + (\del_m r^3)^2 \Big\}
- \half \Big( \frac{H_{\varrho}}{K_{\varrho}} + g^2 H_{\varrho}' \Big)
\Big\{ (\del_m y^2)^2 + (\del_m \wt{\vartheta})^2 \Big\}
\nn \\
\ & \ \ \ \ 
- \Big( \frac{\omega_{\varrho}}{K_{\varrho}} + \omega'_{\varrho} \Big)
\ve^{mn} \, (\del_m y^2) (\del_n \wt{\vartheta}) 
\nn \\
\ & \ \ \ \ 
- \sqrt{2} \, \ve^{mn} \del_m \big( (\vartheta - t_{2}) A_{n} \big)
\, . \label{X8X9-NLSM}
\end{align}
Compared with the string worldsheet sigma model (\ref{string-NLSM}),  
we can read off the target space variables
$G_{IJ}$ and $B_{IJ}$ in the following way (for the indices, see Table \ref{ST-WS}):
\bsubeq \label{G-B-X8X9}
\begin{gather}
G_{66} \ = \ G_{77} \ = \ 
H_{\varrho} + \frac{1}{g^2} H'_{\varrho}
\, , \ls
G_{88} \ = \ G_{99} \ = \
\frac{H_{\varrho}}{K_{\varrho}}
+ g^2 H'_{\varrho} 
\, , \\
B_{89} \ = \ 
- \frac{\omega_{\varrho}}{K_{\varrho}}
- \omega'_{\varrho}
\, . 
\end{gather}
\esubeq
Due to the smearing procedure (\ref{8-isometry}) and (\ref{9-isometry}),
all of these variables do not depend on $y^2$ and $\wt{\vartheta}$. 
The third line in the right-hand side of (\ref{X8X9-NLSM}) is the dyonic mode with the field $\vartheta$, 
which is dual to the dynamical field $\wt{\vartheta}$.
This term also appears in the NLSM (\ref{NLSM-522}).

We note that the configuration (\ref{G-B-X8X9}) does not correspond to (\ref{G-B-522}) given by the NLSM (\ref{NLSM-522}).
This might be interpreted as a deformation of the background geometry of the exotic $5^2_2$-brane.
However, this is quite a naive guess.
We should also notice that 
the variables in (\ref{G-B-X8X9}) do not satisfy 
the equations of motion in supergravity theories. 
The main reason is that the $\N=(4,4)$ supersymmetry is broken down in the presence of $\Scr{L}_{\text{G}}$ (\ref{X8X9-add}).
We will discuss this serious issue in section \ref{summary}. 
Fortunately, 
we can restore $\N=(4,4)$ supersymmetry if we correctly tune the parameters in the sigma model (\ref{X8X9-NLSM}).

\subsection{$\N=(4,4)$ limit}
\label{44-limit} 

The NLSM (\ref{X8X9-NLSM}) has only $\N=(2,2)$ supersymmetry.
This is different from the NLSM for the exotic $5^2_2$-brane with $\N=(4,4)$ supersymmetry (\ref{NLSM-522}).
In order to go back to the NLSM (\ref{NLSM-522}),
we should take a special configuration in which 
the functionals $H'_{\varrho}$ and $\omega'_{\varrho}$ in (\ref{9-isometry}) shrink to zero.
Both variables are proportional to the parameter $\sigma' \sim 1/\wt{\cal R}_9$. 
If we want to consider the vanishing limit of $H'_{\varrho}$ and $\omega'_{\varrho}$, 
we should take the large $\wt{\cal R}_9$ limit.
We now study the background geometry of the exotic $5^2_2$-brane dualized from the geometry of the H-monopole.
In this configuration it is natural to take the large $\wt{\cal R}_9$ limit.
In the same analogy it is also natural to take the small ${\cal R}_8$ limit, 
where the variables $H_{\varrho}$ and $\omega_{\varrho}$ are large.
This limit corresponds to the large $\wt{\cal R}_8$ limit, 
where $\wt{\cal R}_8$ is the radius of the physical coordinate $\wt{X}^8$.
Under this limit, the NLSM (\ref{X8X9-NLSM}) is reduced to
\begin{align}
\Scr{L}_{\text{E2b}}
\ &= \ 
- \frac{1}{2} H_{\varrho} 
\Big\{ (\del_m r^1)^2 + (\del_m r^3)^2 \Big\}
- \frac{H_{\varrho}}{2 K_{\varrho}} 
\Big\{ (\del_m y^2)^2 + (\del_m \wt{\vartheta})^2 \Big\}
\nn \\
\ & \ \ \ \ 
- \frac{\omega_{\varrho}}{K_{\varrho}} 
\ve^{mn} \, (\del_m y^2) (\del_n \wt{\vartheta}) 
- \sqrt{2} \, \ve^{mn} \del_m \big( (\vartheta - t_{2}) A_{n} \big)
\, . \label{X8X9-NLSM-red}
\end{align}
Here the FI parameters $s_1$, $t_1$ and $t_2$ are still arbitrary.
Indeed we can interpret the parameters $(s_1, t_1)$ as the position of the $5^2_2$-brane in the two-dimensional $(r^1, r^3) = (X^6, X^7)$ plane.
The parameter $t_2$ is a shift parameter of the brane along the dual coordinate $\vartheta$. 
Since the NLSM (\ref{NLSM-522}) indicates that the $5^2_2$-brane is located at the origin of the plane,
the NLSM (\ref{X8X9-NLSM}) is a natural generalization of (\ref{NLSM-522}).

\section{Worldsheet instanton corrections}
\label{Instantons}

In this section, we study worldsheet instanton effects in the Remodeled GLSM. 
It is known that string worldsheet instanton effects modify the geometry of the target spacetime \cite{Witten:1993yc}.
The worldsheet instantons in NLSMs are traced back into the gauge
instantons in GLSMs.

Consider an NS5-brane and compactify the transverse direction $X^9$ on $S^1$. We call this $X^9$-circle.
The geometry of the NS5-brane involves infinite tower of the KK-modes in the $X^9$-circle.
When all the KK-modes are smoothed out, the geometry becomes that of the H-monopole.
Then the geometry has an isometry along the $X^9$-circle and one can perform the T-duality transformation using the Buscher rule.
Instantons in the GLSM for the H-monopole are studied in \cite{Tong:2002rq}.
It is shown that the worldsheet instanton effects break the isometry in the geometry.
Then the geometry of the H-monopole becomes that of the NS5-brane on $S^1$ after the worldsheet instanton corrections are involved. 
From the viewpoint of supergravity theories, the disappearance of the isometry is interpreted as the recovery of the light KK modes. 
A physical interpretation of this result is as follows:
In \cite{Tong:2002rq}, the author studied the GLSM in a specific parameter region $g \to 0$ where the radius of $S^1$ shrinks to zero size.
Therefore the KK modes in $S^1$ becomes light and they appear in the string spectrum and modify the geometry.

On the other hand, the H-monopole becomes the KK-monopole by the T-duality transformation along the $X^9$-circle.
The parameter region $g \to 0$ corresponds to the divergent radius of the T-dualized $\wt{X}^9$-circle. 
This indicates that the KK-modes become massive while the winding modes become lighter and appear in the string spectrum. 
Compared with the H-monopole case, the geometry of the KK-monopole should involve information of the light winding modes in this parameter region.
In \cite{Harvey:2005ab} the authors studied the instantons in the GLSM for the KK-monopole. 
They found that the instantons break the isometry along the {\it winding coordinate} $X^9$ and the KK-monopole geometry acquires 
the $X^9$ (not the geometrical coordinate $\widetilde{X}^9$) dependence.
This is consistent with the physical intuition and seems a conceivable result.

The KK-monopole geometry has an isometry along the $\wt{X}^9$-circle. 
The $5^2_2$-brane geometry is obtained by performing the T-duality transformation along the other transverse direction (the $X^8$-direction in this paper) of the KK-monopole.
Then the $5^2_2$-brane geometry has two isometries in the transverse directions $\wt{X}^8$, $\wt{X}^9$.
In the previous papers, we studied the worldsheet instanton corrections to the $5^2_2$-brane geometry through the $\N = (4,4)$ GLSM \cite{Kimura:2013fda, Kimura:2013zva} in the parameter region $g \to 0$.
We found that the $5^2_2$-brane geometry is corrected by instanton effects and the geometry has the $X^9$ winding coordinate dependence.
This is a reflection of the fact that the parameter region $g \to 0$ corresponds to the large radius of the $\wt{X}^9$- (and also $\wt{X}^8$-) circle, and the light winding modes are favored in the spectrum.
However, the $5^2_2$-brane geometry has 
the symmetry under the exchange of the $\wt{X}^8$ and $\wt{X}^9$ directions. 
Therefore it is natural to study the stringy corrections to the $\wt{X}^8$ isometry direction.

In this section we look for the instanton corrections to the second T-dual circle of the $5^2_2$-brane geometry.

\subsection{Truncated model}

Before going to the calculation of the instantons, 
we look for the parameter region where the instanton effects capture the stringy corrections to the geometry. 
We expect that the Remodeled GLSM (\ref{X8X9-GLSM}) introduced in the previous section incorporates
the instanton corrections associated with the $\wt{X}^8$-circle.
In the Remodeled GLSM, there is the topological term $-\sqrt{2} y^{2} F'_{01}$ in the Lagrangian $\Scr{L}_{\text{G}}$.
When the gauge field $A'_{m}$ resides in the non-trivial homotopy class, this term breaks the isometry along $y^2 = \wt{X}^8$ direction. 
Therefore, the instanton configuration in $\Scr{L}_{\text{G}}$ sector can induce the corrections to the geometry caused by the light KK modes.
In order to favor the light KK modes, we take the parameter region $g \to \infty$ where the radii of the $\wt{X}^8$, $\wt{X}^9$-circles become large. 

We first find the truncated model of the Remodeled GLSM in the parameter region $g \to \infty$.
The bosonic part of the Lagrangian is given by (\ref{X8X9-GLSM-b2}) supplemented by the
constraints \eqref{const-add-X8X9-GLSM} and \eqref{const-add2-X8X9-GLSM}.
The supersymmetric vacuum condition is given in (\ref{SUSYvacua-X8X9-GLSM}).
The solutions to the condition
are (\ref{qqt-X8X9-GLSM}) and (\ref{q'q't-X8X9-GLSM}).
We then consider the parameter region $g \to \infty$.
The kinetic terms 
$- \frac{1}{2g^2} [(\del_m r^1)^2 + (\del_m r^3)^2]$ 
in the Lagrangian are dropped out and $r^1$, $r^3$ become auxiliary fields.
From the kinetic terms 
$- \frac{g^2}{2} [(\del_m y^2)^2 + (D_m \gamma^4)^2]$, 
the field $y^2$ is frozen and $D_m \gamma^4 = 0$.
The condition $D_m \gamma^4 = \del_m \gamma^4 + \sqrt{2} A_m = 0$ 
implies that $\gamma^4$ is frozen and $A_m = 0$.
Since $y^2$ is frozen the constraint \eqref{const-add-X8X9-GLSM} in $g \to \infty$ is trivially satisfied. 
The last term in (\ref{X8X9-GLSM-b2}) vanishes by the condition
$\del_m y^2 = 0$
and the constraint \eqref{const-add2-X8X9-GLSM}.
The field $\sigma'$ is decoupled
from the other parts of the Lagrangian and we choose $\sigma' = 0$.
From the potential term 
$- 2 (|\sigma|^2 + 4 |M_c|^2) (|q|^2 + |\wt{q}|^2 + g^2)$, 
the fields $\sigma$ and $M_c$ should stay in the vacuum $\sigma = M_c = 0$ for finite energy configurations.
Then the bosonic part of the truncated model for the Remodeled GLSM in the limit $g \to \infty$ is 
\begin{align}
\Scr{L}_{\text{t}} 
\ &= \ 
\Scr{L}_1 + \Scr{L}_2
\, , 
\end{align}
where 
\bsubeq
\begin{align}
\Scr{L}_1 \ &= \ 
- \big\{  (\del_m q)^2 + (\del_m \wt{q})^2 \big\} 
- \frac{1}{e'{}^2} |\del_m \phi'|^2 
- 2 |\phi'|^2 (|q|^2 + |\wt{q}|^2) 
\nn \\ 
\ & \ \ \ \  
- \frac{e^2}{2} 
\Big\{ |q|^2 - |\wt{q}|^2 - \sqrt{2} (r^3 - t_1) \Big\}^2
- e^2 
\Big|
\sqrt{2} q \wt{q} - \big( (r^1 - s_1) + \i (r^2 - s_2) \big)
\Big|^2
\, , \\
\Scr{L}_2 
\ &= \ 
\frac{1}{2 e'{}^2}
(F'_{01})^2
- \Big\{ |D_m q'|^2 + |D_m \wt{q}'|^2 \Big\}
- \sqrt{2} (y^2 - t'_2) F'_{01} 
\nn \\
\ & \ \ \ \ 
- e'{}^2 \Big|
\sqrt{2} \, q' \wt{q}' - \big( \i \gamma^4 - \sqrt{2} s' \big)
\Big|^2
\, .
\label{eq:truncated1}
\end{align}
\esubeq
We note that $\Scr{L}_1$ and $\Scr{L}_2$ are completely independent of each other.
We also stress that the gauge field $A_m$, hence the associated topological term, is dropped in the $g \to \infty$ truncated model.
This is the most notable difference from the $g \to 0$ truncated model discussed in \cite{Kimura:2013zva} where the $A_m$ gauge dynamics induces the $X^9$ winding mode corrections through the $A_m$ topological term. 
Although the model \eqref{eq:truncated1} fails to capture the $X^9$ winding mode corrections, it can potentially break the isometry along the $y^2$ direction by the $A_m'$ topological term. 
In order to find the field configuration of the gauge field $A'_m$ with non-trivial homotopy class, we further focus on a specific field configuration.

Since the dynamics of $\Scr{L}_1$ and $\Scr{L}_2$ is independent, 
we consider the $\Scr{L}_2$ part in the following.
The field $\gamma^4$ is frozen and should be treated as a vacuum moduli $\xi \in \mathbb{R}$.
For simplicity, we consider a specific configuration 
$q' = \frac{1}{\sqrt{2}} f$, $\wt{q}' = \frac{\i}{\sqrt{2}} \ol{f}$
where $f$ is a complex scalar field.
Then we find 
\begin{align}
\Scr{L}_2' 
\ &= \
\frac{1}{2 e'{}^2} (F'_{01})^2 
- |D_m f|^2 - \frac{e'{}^2}{2} \left( |f|^2 - \sqrt{2} \xi \right)^2 
- \sqrt{2} (y^2 - t'_2) F'_{01}
\, ,
\end{align}
where we have turned on only the imaginary part of $s'$ 
which is absorbed into the vacuum moduli $\xi$. 
The resultant model $\Scr{L}_2'$ is nothing but the Abelian-Higgs model in two dimensions. 
The instantons in this gauge field theory are known as vortices. 
In the following subsection, 
we calculate instanton corrections to the target spacetime geometry in the IR regime.

\subsection{BPS vortices}

After the Wick rotation to the Euclidean space, the Lagrangian $\Scr{L}'_2$ is rewritten as  
\begin{align}
\Scr{L}'_2
\ &= \  
\frac{1}{2e'{}^2} (F'_{12})^2 
+ |D_m f|^2 
+ \frac{e'{}^2}{2} 
\left(
|f|^2 - \sqrt{2} \xi \right)^2 
+ \i \sqrt{2} (y^2 - t'_2) F'_{12} 
\nn \\
\ &= \ 
\frac{1}{2 e'{}^2}
\Big\{
F'_{12} \pm e^{\prime 2} \big( |f|^2 - \sqrt{2} \xi \big)
\Big\}^2
+ \big| (D_1 \pm \i D_2) f \big|^2 
\pm \sqrt{2} \xi F'_{12} 
+ \i \sqrt{2} (y^2 - t'_2) F'_{12}
\, .
\end{align}
The instantons in the model are just the Abrikosov-Nielsen-Olesen (ANO) vortices.
The BPS vortex equations are 
\begin{align}
F'_{12} \pm e'{}^2 \left( |f|^2 - \sqrt{2} \xi \right) \ = \ 0
\, , \ls
(D_1 \pm \i D_2) f \ = \ 0
\, . 
\label{eq:BPSeq}
\end{align}
In order that the BPS state has finite energy, the field $|f|^2$ should
asymptotics to the vacuum $\sqrt{2} \xi$.
This is indeed the case\footnote{In this case, the vacuum of the $q'$, $\wt{q}'$ is given by
$q' = - \frac{\i}{2^{1/4}} e^{- \i \alpha'} \sqrt{ \gamma^4 - s'_2}$,
$\wt{q}' = - \frac{1}{2^{1/4}} e^{\i \alpha'} \sqrt{\gamma^4 - s'_2}$.
} when we take $t_1' = 0$.
Then the action associated with $\Scr{L}'_2$ is evaluated in this
BPS state as 
\begin{align}
S'_{2} \ &= \ 
\pm \sqrt{2} \xi \int \d^2 x \, F'_{12} 
+ \i \sqrt{2} \, (y^2 - t'_{2}) \int \! \d^2 x \, F'_{12}
\, . \label{eq:inst_action}
\end{align}

Now we study the instanton corrections to the target spacetime geometry. 
The analysis attributes to the four-point functions of fermions
$\psi'_{\pm}, \wt{\psi}'_{\pm}$ in the charged hypermultiplet $(Q', \wt{Q}')$ in the background of the vortex solutions \eqref{eq:BPSeq}.
Each fermion $\psi'_{\pm}$, $\wt{\psi}'_{\pm}$ contains one Goldstino mode and four fields are enough to saturate these fermionic moduli integral.
The other fermionic moduli are saturated by the instanton moduli action including the Riemann tensor in the moduli space \cite{Tong:2002rq}.
The calculations of the four-point function are the same performed in \cite{Tong:2002rq, Harvey:2005ab, Kimura:2013zva} and we never repeat it here.

In the string worldsheet sigma model (\ref{string-NLSM}), the fermions $\psi'_{\pm}, \wt{\psi}'_{\pm}$ are related to the superpartners of the geometrical coordinate fields via the supersymmetric completion of the vacuum 
condition \eqref{constq'q'}. 
The four-point interaction of the fermions associated with the geometrical coordinates is interpreted as the Riemann tensor of the target spacetime geometry. Therefore the Riemann tensor receives instanton effects from which we find the corrections to the geometry.
The relations among the fermions are derived from the superfield equations of motion in the IR limit.
In the IR limit $e' \to \infty$, the superfield Lagrangian of the charged hypermultiplet is given by 
\begin{align}
\Scr{L}^{\text{IR}}_{\text{G}} 
\ &= \  
\int \! \d^4 \theta \,
\Big\{
\ol{Q}{}' \e^{-2V'} Q' + \ol{\wt{Q}} \e^{+2V'} \wt{Q}' \Big\}
 + \Big\{ \sqrt{2} \int \! \d^2 \theta \ 
\big( \wt{Q}' \Phi' Q' + (s' - \Gamma) \Phi' \big)
+ \text{(h.c.)} \Big\}
\nn \\
\ & \ \ \ \ 
+ \Big\{ \sqrt{2} \int \! \d^2 \wt{\theta} \, 
(t' - \Xi) \Sigma' + \text{(h.c.)} \Big\}
\, .
\end{align}
The kinetic term of the vector multiplet $(V', \Phi')$ is dropped out
in the IR limit and they become auxiliary fields.
The twisted F-term is rewritten as a D-term and a total derivative term:
\begin{align}
\sqrt{2} \int \! \d^2 \wt{\theta} \, 
(t' - \Xi) \Sigma' + \text{(h.c.)} 
\ &= \  
- 2 \int \! \d^4 \theta \, 
(\Xi + \ol{\Xi})  V' 
- \sqrt{2} \, \ve^{mn} \, \del_m ((y^2 - t'_2) A'_n)
\, .
\end{align}
Then the equations of motion for $V'$, $\Phi'$ are
\bsubeq \label{eq:super_eom} 
\begin{align}
0 \ &= \ 
\ol{Q}{}' \e^{-2V'} Q' - \ol{\wt{Q}}{}' \e^{+2V'} \wt{Q}' 
- (\Xi + \ol{\Xi})
\, , \\
0 \ &= \ 
\wt{Q}' Q' + (s'-\Gamma) 
\, .
\end{align}
\esubeq
From which we find the constraints among the charged hyper fermions and
the superpartner of the geometrical coordinates $(r^1, r^3, y^2, \gamma^4)$:
\bsubeq
\begin{align}
\frac{1}{g^2} \chi_{\pm} - \sqrt{2} (\psi_{c\pm} + \ol{\chi}_{c \pm})
\ &= \ 
- (\ol{q}{}' \psi'_{\pm} - \ol{\wt{q}}{}' \wt{\psi}'_{\pm})
\, , \\
\frac{1}{g^2} \wt{\chi}_{\pm} 
\ &= \
- (\ol{\wt{q}}{}'
 \ol{\psi}{}'_{\pm} + \ol{q}' \ol{\wt{\psi}}{}'_{\pm})
\, .
\end{align}
\esubeq
In the vacuum, we choose $\psi_{c \pm} = \chi_{c\pm} = 0$.
Then we have  
\begin{align}
\psi'_{\pm} 
\ &= \ 
- \frac{f}{\sqrt{2} g^2 |f|^2} (\chi_{\pm} - \i \ol{\wt{\chi}}_{\pm})
\, , \ls
\wt{\psi}'_{\pm} 
\ = \
\frac{\i \ol{f}}{\sqrt{2} g^2 |f|^2}
 (\chi_{\pm} + \i \ol{\wt{\chi}}_{\pm})
\, , 
\end{align}
where 
\begin{align}
|f|^2 \ = \ |q'|^2 + |\wt{q}'|^2 \ = \ \sqrt{2} \xi
\, .
\end{align}
The four-point function which receives instanton corrections is \cite{Tong:2002rq, Harvey:2005ab} 
\begin{align}
G^{(k)} \ &= \ 
\left. \langle 
\ol{\psi}{}'_{+} (x_1) \psi'_{-} (x_2) \wt{\psi}'_{+} (x_3)
 \ol{\wt{\psi}}{}'_{-} (x_4) 
\rangle 
\right|_{\text{$k$-inst.}}
\, ,
\label{eq:4ptfunc}
\end{align}
where the correlation function is evaluated in the instanton background with topological number $k$.
This provides the four-point interaction term 
$\ol{\psi}{}'_{+} \psi'_{-} \wt{\psi}'_{+} \ol{\wt{\psi}}{}'_{-}$ in the IR Lagrangian.

In order to interpret the four-point interaction of $\psi'_{\pm}$,
$\wt{\psi}'_{\pm}$ in terms of the geometrical fermions, 
we need to identify the superpartners of 
real components $(r^1, r^3, y^2, \gamma^4)$ in the superfields $(\Xi, \Gamma)$.
Those are extracted from the real and imaginary parts of the superfields $(\Xi, \Gamma)$, namely, 
\bsubeq
\begin{alignat}{2}
\Xi_1 \ &= \ 
\frac{1}{\sqrt{2}} (\Xi + \ol{\Xi})
\, , &\ls
\Xi_2 \ &= \
- \frac{\i}{\sqrt{2}} (\Xi - \ol{\Xi})
\, , \\
\Gamma_1 \ &= \
\frac{1}{\sqrt{2}} (\Gamma + \ol{\Gamma})
\, , &\ls
\Gamma_2 \ &= \
- \frac{\i}{\sqrt{2}} (\Gamma - \ol{\Gamma})
\, .
\end{alignat}
\esubeq
They are compared with an $\N = (1,1)$ real superfield 
$\ol{R} = R = A + \sqrt{2} \theta^{+}_r \Omega_{+} + \sqrt{2} \theta^{-}_r \Omega_{-} + \ldots$,
where $\theta^{\pm}_r$ are
real fermionic coordinates. 
We first decompose the $\N = (2,2)$ coordinates $\theta^{\pm}$
into the real and imaginary components 
$\theta^{\pm} = \theta^{\pm}_R + \i \theta^{\pm}_I$ where $\theta^{\pm}_R$ and $\theta^{\pm}_I$ are real. 
We follow the convention of \cite{Harvey:2005ab} where the $\N = (1,1)$ 
coordinates are the imaginary parts\footnote{For an $\N = (2,2)$ twisted superfield, we need to flip the sign of $\theta^{-}_I$ to define the real superfield.} of $\theta^{\pm}$.
Then we find that the real fermions 
$(\Omega^{r^1}, \Omega^{r^3}, \Omega^{y^2}, \Omega^{\gamma^4})$ associated with the geometrical coordinates $(r^1, r^3, y^2, \gamma^4)$ are 
\bsubeq
\begin{alignat}{2}
\Omega^{r^1}_{\pm} \ &= \ 
- (\chi_{\pm} - \ol{\chi}_{\pm})
\, , &\ls
\Omega^{r^3}_{\pm} \ &= \ 
\wt{\chi}_{\pm} - \ol{\wt{\chi}}_{\pm}
\,, \\
\Omega^{y^2}_{\pm} \ &= \ 
- \frac{\i}{g^2} (\chi_{\pm} + \ol{\chi}_{\pm})
\, , &\ls
\Omega^{\gamma^4}_{\pm} \ &= \ 
- \frac{\i}{g^2} (\wt{\chi}_{\pm} + \ol{\wt{\chi}}_{\pm})
\, .
\end{alignat}
\esubeq
From this expression we conclude that terms containing
$\Omega^{y^2}$ and $\Omega^{\gamma^4}$ in the four-point interaction \eqref{eq:4ptfunc} can survive in the limit $g \to \infty$. 
The genuine geometrical coordinate is not $\gamma^4$ but $\wt{\vartheta} = \gamma^4 + \sqrt{2} \alpha$, 
where $\alpha$ is the phase of $q$, $\wt{q}$.
The superpartner of $\alpha$ is found in \cite{Harvey:2005ab} but this is irrelevant in the $g \to \infty$ limit, 
and we can take $\Omega^{\wt{\vartheta}} = \Omega^{\gamma^4}$.
Then in the vacuum and the limit $g^2 \to \infty$, we find 
\begin{align}
\psi'_{\pm} \ &= \ 
\frac{-\i f}{2 \sqrt{2} |f|^2}
(\Omega^{y^2}_{\pm} - \i \Omega^{\wt{\vartheta}}_{\pm})
\, , \ls
\wt{\psi}'_{\pm} \ = \ 
\frac{\ol{f}}{2 \sqrt{2} |f|^2} 
(\Omega^{y^2}_{\pm} + \i \Omega^{\wt{\vartheta}}_{\pm})
\, .
\end{align}
Using this expression, we find that the instanton corrections to the four-point interaction 
$\ol{\psi}{}'_{+} \psi'_{-} \wt{\psi}'_{+} \ol{\wt{\psi}}{}'_{-}$ vanish.
This result is quite different from that in the
parameter region $g \to 0$ where the instanton corrections to the four-point interaction remain non-zero.
Although the parameter region $g \to \infty$ suggests that the radius of
the $\widetilde{X}^8$-circle becomes large, the Remodeled GLSM in $g \to
\infty$ does not capture the light KK-modes which would appear in the
spectrum. We give a short discussion for this result in the next section.

\section{Summary and discussions}
\label{summary}

In this paper, we remodeled the $\N=(4,4)$ supersymmetric GLSM for the exotic $5^2_2$-brane \cite{Kimura:2013fda, Kimura:2013zva}.
We added another vector multiplet $(V', \Phi')$, another charged
hypermultiplet $(Q', \wt{Q}')$ and other FI parameters $(s', t')$.
Coupled them to the neutral hypermultiplet $(\Xi, \Gamma)$ in additional (twisted) F-terms, 
we constructed a new model (\ref{X8X9-GLSM}), called the Remodeled GLSM.
This is a sigma model whose target space geometry has two gauged isometries.
The Remodeled GLSM has only $\N=(2,2)$ supersymmetry rather than $\N=(4,4)$ supersymmetry.
This is caused by a conflict between two different $SU(2)_R$ symmetries associated with two respective vector multiplets.
The IR effective theory is also affected by this conflict.
However, if we choose a suitable choice of parameters in the IR limit,
i.e., if we take the small $\sigma'$ limit,
we can restore the NLSM of the background geometry of the single exotic $5^2_2$-brane.
We can also interpret that the $\N=(2,2)$ supersymmetry is enhanced to $\N=(4,4)$ supersymmetry.
We also pointed that the original background geometry is realized in the small $\sigma'$ limit.

We also studied the quantum corrections to the background geometry of the exotic $5^2_2$-brane from the Remodeled GLSM (\ref{X8X9-GLSM}).
Since the GLSM can be regarded as the UV completion of the string worldsheet sigma model,
the string worldsheet instanton corrections can be traced by the vortex corrections.
This technique is quite successful to investigate the quantum corrections to the H-monopoles and to the KK-monopoles \cite{Tong:2002rq, Harvey:2005ab, Okuyama:2005gx}.
Previously, we also applied this to the GLSM with one gauged isometry \cite{Kimura:2013zva}. 
There the isometry along the $X^9$-circle is gauged.
We could argued the quantum corrections to the $X^9$-circle of the $5^2_2$-brane.
We understood that the corrections are generated by the string winding modes to the dual coordinate $X^9$.
The Remodeled GLSM (\ref{X8X9-GLSM}) has two gauged isometries
along not only the $X^9$-circle but also the $\wt{X}^8$-circle.
We studied the Remodeled GLSM in the parameter region $g \to \infty$
where the radii of the $\wt{X}^8$- and $\wt{X}^9$-circles become large. 
The $g \to \infty$ limit allows us to truncate the model, in which several fields are frozen and lose their dynamics.
Then the Remodeled GLSM is reduced to the Abelian-Higgs model with the decoupled sector.
We showed that the Abelian-Higgs model accommodates the instanton solution. The instanton corrections to the four-point correlation function of the charged fermions are calculated in the standard way.
In the IR limit, the four-point interaction is interpreted as that of the superpartner of the geometrical coordinate. We found that the four-point interaction of the geometrical fermions vanishes in the limit $g \to \infty$ and is not captured in the Remodeled GLSM.


As we summarized above,
we investigated the Remodeled GLSM (\ref{X8X9-GLSM}) in the classical and quantum levels.
We found various issues which we have to solve in a near future.
Here we enumerate two of them with our current interpretations.

\subsection*{Multiple defect five-branes?}

We have constructed the Remodeled GLSM (\ref{X8X9-GLSM}) as the supersymmetric sigma model with two gauged isometries. 
Indeed, in the large $\wt{\cal R}_9$ limit, 
we derived the NLSM of the background geometry of the exotic $5^2_2$-brane.
This is identical to the NLSM (\ref{NLSM-522}).
Furthermore, we wonder whether this NLSM itself would have much wider feature of five-branes, 
even though the target space variables (\ref{G-B-X8X9}) do not satisfy 
field equations of supergravity theories.

Let us remember the GLSMs with multiple $U(1)$ vector multiplets (\ref{GLSM-HM}) and (\ref{GLSM-KKM}), which have been discussed in \cite{Okuyama:2005gx}.
In each model we prepare $k$ sets of the vector multiplets, the FI parameters, and the charged hypermultiplets.
In the IR limit, the $k$ sets yield $k$ five-branes.
The gauge coupling of each vector multiplet in the (twisted) F-term and the D-term determines the configuration of each five-brane as an H-monopole (i.e., a smeared NS5-brane) or a KK-monopole.
Each set of the FI parameters indicates the position of each five-brane.

We again consider the Remodeled GLSM (\ref{X8X9-GLSM}).
This GLSM has two vector multiplets $(V, \Phi)$ and $(V', \Phi')$.
The former vector multiplet is coupled to the neutral hypermultiplet $(\Xi, \Gamma)$ in the D-terms. 
This generates the exotic $5^2_2$-brane (\ref{GLSM-522}).
The latter vector multiplet is coupled to $(\Xi, \Gamma)$ in the twisted F-term and in the F-term.
This coupling is the same as the one to generate the H-monopole (\ref{GLSM-HM}), 
as we mentioned before.
If we apply the above discussion of the GLSM for multiple five-branes to the Remodeled GLSM (\ref{X8X9-GLSM}),
we encounter the following speculations:
\begin{itemize}
\item Does the Remodeled GLSM (\ref{X8X9-GLSM}) provide the two-body system of an exotic $5^2_2$-brane and a defect NS5-brane in the IR limit?
\item Do the FI parameters $(s,t)$ and $(s',t')$ represent the positions of the exotic $5^2_2$-brane and the defect NS5-brane, respectively?
\end{itemize}
We would be able to argue these speculations partially,
even though it is hard to understand them completely in the current stage.
Focusing on the parameters $\sigma \sim 1/\wt{\cal R}_9$ and $\sigma' \sim 1/{\cal R}_8$ in (\ref{8-isometry}) and (\ref{9-isometry}), 
we consider the following two configurations:
\begin{enumerate}
\item {a configuration that an exotic $5^2_2$-brane is dominant: $\sigma' \to 0$ and $\sigma \to \infty$}

This is the case which we have already studied in section \ref{44-limit}.
In this limit the functionals $(H'_{\varrho}, \omega'_{\varrho})$ originated from the vector multiplet $(V', \Phi')$, the FI parameters $(s', t')$ and the charged hypermultiplet $(Q', \wt{Q}')$ are removed.
We obtain the NLSM for the background geometry of an exotic $5^2_2$-brane.

\item {a configuration that a defect NS5-brane is dominant: $\sigma \to 0, \sigma' \to \infty$ and $g \to 1$}

This is another remarkable limit.
In this configuration, 
the functionals $(H_{\varrho}, \omega_{\varrho})$ are reduced to simple values, 
whilst $(H'_{\varrho}, \omega'_{\varrho})$ are unchanged.
The NLSM (\ref{X8X9-NLSM}) is eventually reduced to the following form\footnote{The $g \to 1$ limit is necessary to yield the geometry of the defect NS5-brane.}:
\bsubeq \label{multiview-X8X9-NS5}
\begin{align}
H_{\varrho} \ &\to \ 1
\, , \ls
\omega_{\varrho} \ \to \ 0
\, , \ls
\wh{H}_{\varrho} \ = \ 
1 + H'_{\varrho}
\ = \ 
1 + \sigma' \log \frac{\Lambda'}{\varrho'} 
\, , \\
\Scr{L}_{\text{E2b}}
\ &= \ 
- \frac{1}{2} \wh{H}_{\varrho} \Big\{ (\del_m r^1)^2 + (\del_m r^3)^2 
+ (\del_m y^2)^2 + (\del_m \wt{\vartheta})^2 \Big\}
\nn \\
\ & \ \ \ \ 
- \omega'_{\varrho} \, \ve^{mn} \, (\del_m y^2) (\del_n \wt{\vartheta}) 
- \sqrt{2} \, \ve^{mn} \del_m \big( (\vartheta - t_{2}) A_{n} \big)
\, .
\end{align}
\esubeq
This is nothing but the NLSM of the background geometry of a defect NS5-brane (for a similar configuration, see also appendix \ref{app-H2}).
The position of the defect NS5-brane is given by the FI parameters 
$(- t'_1, - s'_1)$ in the $(r^1, r^3) = (X^6, X^7)$ plane,
which can be read from the functional $\varrho'$ in (\ref{9-isometry}).
\end{enumerate}

Due to the above two configurations,
we suspect that the target space geometry of the NLSM (\ref{X8X9-NLSM}) would describe ``a part of'' the two-body system of an exotic $5^2_2$-brane and a defect NS5-brane.
Unfortunately, however, we have not obtained the perfect construction of the two-body system yet. 
This is because the target space geometry (\ref{G-B-X8X9}) does not
satisfy the field equations in supergravity theories. 
We cannot find a suitable solution of the dilaton field, either.

\subsection*{Worldsheet instanton effects}

One may expect that the Remodeled GLSM in the parameter region $g \to \infty$ could capture the light KK-modes and modify the geometry.
However, we found that the instantons from the second gauge field 
in the Remodeled GLSM (\ref{X8X9-GLSM}) do not induce any corrections to the geometry in the limit $g \to \infty$.
The resolution to this puzzle lies in the fact that the Remodeled GLSM constructed in this paper may not be enough to describe the whole physics of five-branes.
We think that the Remodeled GLSM (\ref{X8X9-GLSM}) represents not only the sigma model for a single defect five-brane with two gauged isometries,
but also ``a part of'' the two-body system of an exotic $5^2_2$-brane and a defect NS5-brane as we have already argued above.
The gauge multiplet $(V', \Phi')$ is associated with the geometry of the H-monopole compactified on the $\wt{X}^8$-circle (the defect NS5-brane) in the IR limit.
When the gauge field $A'_m$ resides in the non-trivial topological sector, the configuration breaks the isometry along the $\wt{X}^8$-circle.
Then the instantons would uplift the defect NS5-brane of codimension two to the H-monopole of codimension three.
This is the reminiscent of the instanton corrections to the geometry of the H-monopole studied in \cite{Tong:2002rq}.
The NS5-brane of codimension four is recovered from the H-monopole of codimension three by the instanton effects. 
In this paper we showed that the instantons in the limit $g \to \infty$ live in the gauge multiplet $(V', \Phi')$ in the truncated Lagrangian $\Scr{L}_2$.
However, the defect NS5-brane appears in the specific limit of the parameter $g \to 1$.
Therefore the natural geometrical interpretation of the instanton effects contradicts with the truncated model. 


\vspace{5mm}

In order to find the complete description both in the classical and in the quantum levels, we have to find additional terms to the Remodeled GLSM (\ref{X8X9-GLSM}).
Since we have not obtained any suitable ideas in the current stage,
we should bear in mind the above two discussions as future problems.
If we successfully construct a GLSM which solves them, 
we can consistently describes the two-body system.
Furthermore we can apply it to various situations discussed in \cite{deBoer:2012ma} 
and try to analyze nongeometric microstates of black holes in string theory.

\section*{Acknowledgements}

The authors thank
Masaki Shigemori 
and
Satoshi Yamaguchi
for helpful discussions and comments.
They also thank the Yukawa Institute for Theoretical Physics at Kyoto University. Discussions during the YITP workshop on ``Field Theory and String Theory'' ({YITP-W-13-12}) were useful to complete this work. 
The work of TK is supported in part by the MEXT-Supported Program for the Strategic Research Foundation at Private Universities 
from MEXT (Ministry of Education, Culture, Sport, Science and Technology) of Japan (S0901029: Research Center for Measurement in Advanced Science, Rikkyo University).
The work of SS is supported in part by Sasakawa Scientific Research Grant from
The Japan Science Society and Kitasato University Research Grant for
Young Researchers.

\begin{appendix}
\section*{Appendix}

\section{Conventions}
\label{app-SF}

In this appendix 
we summarize the conventions of $\N=(2,2)$ supersymmetry in two-dimensional spacetime.
$\N=(2,2)$ supersymmetric objects are also applicable to describe 
$\N=(4,4)$ supersymmetric theories as discussed in \cite{Tong:2002rq}.

First of all,
we introduce the $\N=(2,2)$ supercovariant derivatives defined as
\begin{align}
D_{\pm} \ &= \ 
\frac{\del}{\del \theta^{\pm}} 
- \i \ol{\theta}{}^{\pm} \big( \del_0 \pm \del_1 \big)
\, , \ls
\ol{D}{}_{\pm} \ = \ 
- \frac{\del}{\del \ol{\theta}{}^{\pm}} 
+ \i \theta^{\pm} \big( \del_0 \pm \del_1 \big)
\, . \label{SCD}
\end{align}
Here $\theta^{\pm}$ are the anti-commuting Grassmann coordinates in superspace.
All the $\N=(2,2)$ superfields are defined in terms of the supercovariant derivatives.
It is also useful to define the integral measures of the Grassmann coordinates such as
\begin{gather}
\d^2 \theta \ = \ 
- \half \, \d \theta^+ \, \d \theta^- 
\, , \ls
\d^2 \wt{\theta} \ = \ 
- \half \, \d \theta^+ \, \d \ol{\theta}{}^- 
\, , \ls
\d^4 \theta 
\ = \ 
- \frac{1}{4} \d \theta^+ \, \d \theta^- \, \d \ol{\theta}{}^+ \,
\d \ol{\theta}{}^- 
\, . \label{f-measure-22}
\end{gather}
The first measure is for F-terms, the second one is for twisted F-terms,
and the third one is for D-terms in the supersymmetric Lagrangian in the $\N=(2,2)$ language.
It is important to describe them explicitly 
in order to avoid any ambiguities caused by the ordering of the Grassmann coordinates. 

As discussed in \cite{Tong:2002rq}, 
the $\N=(4,4)$ string worldsheet sigma model of the background geometry of the NS5-branes
is described by $\N=(2,2)$ supersymmetric objects with $SU(2)_R$ symmetry.
The sigma model of the background geometry of the exotic $5^2_2$-brane,
which emerges by {\rm T}-duality transformations from the background geometry of the NS5-branes,
is also formulated in the language of $\N=(2,2)$ superfields \cite{Kimura:2013fda, Kimura:2013zva}.
Here we exhibit the expansions of the $\N=(2,2)$ superfields.

An $\N=(4,4)$ vector multiplet is built with 
an $\N=(2,2)$ vector superfield $V_a$ 
and an $\N=(2,2)$ chiral superfield $\Phi_a$.
Their expansions by component fields under the Wess-Zumino gauge are
\bsubeq
\begin{align}
V_a \ &= \ 
\theta^+ \ol{\theta}{}^+ (A_{0,a} + A_{1,a})
+ \theta^- \ol{\theta}{}^- (A_{0,a} - A_{1,a})
- \sqrt{2} \, \theta^- \ol{\theta}{}^+ \sigma_a
- \sqrt{2} \, \theta^+ \ol{\theta}{}^- \ol{\sigma}{}_a
\nn \\
\ & \ \ \ \ 
- 2 \i \, \theta^+ \theta^-
\big( \ol{\theta}{}^+ \ol{\lambda}{}_{+,a} 
+ \ol{\theta}{}^- \ol{\lambda}{}_{-,a} \big)
+ 2 \i \, \ol{\theta}{}^+ \ol{\theta}{}^-
\big( \theta^+ \lambda_{+,a} + \theta^- \lambda_{-,a} \big)
- 2 \, \theta^+ \theta^- \ol{\theta}{}^+ \ol{\theta}{}^- D_{V,a}
\, , \\
\Phi_a \ &= \ 
\phi_a 
+ \i \sqrt{2} \, \theta^+ \wt{\lambda}_{+,a} 
+ \i \sqrt{2} \, \theta^- \wt{\lambda}_{-,a}
+ 2 \i \, \theta^+ \theta^- D_{\Phi,a}
+ \dots
\, ,
\end{align}
\esubeq
where the term ``$\ldots$'' comes from the derivative expansions by the supercovariant derivatives (\ref{SCD}).
Note that $V_a$ can be written as 
a twisted chiral superfield
as $\Sigma_a = \frac{1}{\sqrt{2}} \ol{D}{}_+ D_- V_a$,
and $\Phi_a$ can be described in terms of (\ref{SCD}) and 
an unconstrained complex superfield
$C_a$ in such a way as $\Phi_a = \ol{D}{}_+ \ol{D}{}_- C_a$.
Their expansions are 
\bsubeq
\begin{align}
\Sigma_a 
\ &= \ 
\sigma_a 
- \i \sqrt{2} \, \theta^+ \ol{\lambda}{}_{+,a} 
- \i \sqrt{2} \, \ol{\theta}{}^- \lambda_{-,a}
+ \sqrt{2} \, \theta^+ \ol{\theta}{}^- (D_{V,a} - \i F_{01,a})
+ \ldots
\, , \\
C_a \ &= \ 
\phi_{c,a} + \i \sqrt{2} \, \theta^+ \psi_{c+,a} 
+ \i \sqrt{2} \, \theta^- \psi_{c-,a} 
+ \i \sqrt{2} \, \ol{\theta}{}^+ \chi_{c+,a} 
+ \i \sqrt{2} \, \ol{\theta}{}^- \chi_{c-,a}
\nn \\
\ & \ \ \ \ 
+ 2 \i \, \theta^+ \theta^- F_{c,a} 
+ 2 \i \, \ol{\theta}{}^+ \ol{\theta}{}^- M_{c,a}
+ 2 \i \, \theta^+ \ol{\theta}{}^- G_{c,a} 
+ 2 \i \, \ol{\theta}{}^+ \theta^- N_{c,a}
+ \theta^- \ol{\theta}{}^- A_{c=,a}
+ \theta^+ \ol{\theta}{}^+ B_{c\+,a}
\nn \\
\ & \ \ \ \ 
- 2 \i \, \theta^+ \theta^- \ol{\theta}{}^+ \zeta_{c+,a}
- 2 \i \, \theta^+ \theta^- \ol{\theta}{}^- \zeta_{c-,a}
+ 2 \i \, \ol{\theta}{}^+ \ol{\theta}{}^- \theta^+ \lambda_{c+,a}
+ 2 \i \, \ol{\theta}{}^+ \ol{\theta}{}^- \theta^- \lambda_{c-,a}
\nn \\
\ & \ \ \ \ 
- 2 \theta^+ \theta^- \ol{\theta}{}^+ \ol{\theta}{}^- D_{c,a}
\, .
\end{align}
\esubeq
The relation
among the component fields of $\Phi_a$ and $C_a$ are
\bsubeq \label{Phi2C}
\begin{align}
\phi_a
\ &= \ 
- 2 \i \, M_{c,a} 
\, , \\
D_{\Phi,a}
\ &= \ 
- \i \, D_{c,a}
+ \frac{1}{2} (\del_0 - \del_1) B_{c\+,a}
- \frac{\i}{2} (\del_0 + \del_1) A_{c=,a}
+ \frac{\i}{2} (\del_0^2 - \del_1^2) \phi_{c,a}
\, , \\
\wt{\lambda}_{\pm,a}
\ &= \ 
- \sqrt{2} \, \lambda_{c\pm,a}
\mp \i (\del_0 \pm \del_1) \chi_{c\mp,a}
\, . 
\end{align}
\esubeq
An $\N=(4,4)$ charged hypermultiplet is
expressed by $\N=(2,2)$ chiral superfields $Q_a$ and $\wt{Q}_a$ whose expansions are
\bsubeq
\begin{align}
Q_a \ &= \ 
q_a 
+ \i \sqrt{2} \, \theta^+ \psi_{+,a} 
+ \i \sqrt{2} \, \theta^- \psi_{-,a}
+ 2 \i \, \theta^+ \theta^- F_a
+ \ldots
\, , \\
\wt{Q}_a \ &= \ 
\wt{q}_a
+ \i \sqrt{2} \, \theta^+ \wt{\psi}_{+,a}
+ \i \sqrt{2} \, \theta^- \wt{\psi}_{-,a}
+ 2 \i \, \theta^+ \theta^- \wt{F}_a
+ \ldots
\, . 
\end{align}
\esubeq
The target space coordinates of the string worldsheet sigma model are provided by 
an $\N=(4,4)$ neutral hypermultiplet whose building blocks are $\N=(2,2)$ chiral and twisted chiral superfields.
In the case of the background geometry of the NS5-branes,
the pair $(\Psi, \Theta)$ is the $\N=(4,4)$ neutral hypermultiplet,
whilst the geometry of the KK-monopoles is given by the pair $(\Psi, \Gamma)$.
The geometry of the exotic $5^2_2$-brane is represented by the pair $(\Xi, \Gamma)$.
Their explicit expansions are
\bsubeq
\begin{align}
\Psi \ &= \ 
\frac{1}{\sqrt{2}} (r^1 + \i r^2)
+ \i \sqrt{2} \, \theta^+ \chi_+
+ \i \sqrt{2} \, \theta^- \chi_-
+ 2 \i \, \theta^+ \theta^- G
+ \ldots
\, , \\
\Theta \ &= \ 
\frac{1}{\sqrt{2}} (r^3 + \i \vartheta)
+ \i \sqrt{2} \, \theta^+ \ol{\wt{\chi}}{}_+
+ \i \sqrt{2} \, \ol{\theta}{}^- \wt{\chi}_-
+ 2 \i \, \theta^+ \ol{\theta}{}^- \wt{G}
+ \ldots
\, , \\
\Gamma \ &= \ 
\frac{1}{\sqrt{2}} (\gamma^3 + \i \gamma^4)
+ \i \sqrt{2} \, \theta^+ \zeta_+
+ \i \sqrt{2} \, \theta^- \zeta_-
+ 2 \i \, \theta^+ \theta^- G_{\Gamma}
+ \ldots
\, , \\
\Xi \ &= \ 
\frac{1}{\sqrt{2}} (y^1 + \i \, y^2)
+ \i \sqrt{2} \, \theta^+ \ol{\xi}{}_+
+ \i \sqrt{2} \, \ol{\theta}{}^- \xi_-
+ 2 \i \, \theta^+ \ol{\theta}{}^- G_{\Xi}
+ \ldots
\, .
\end{align}
\esubeq
The relations among them under the duality transformations \cite{Rocek:1991ps} are given in (\ref{Theta2Gamma}) and (\ref{Psi2Xi}).
As mentioned in the main part of the present paper, 
$\gamma^4$ is rewritten as $\wt{\vartheta}$ under the gauge transformation \cite{Tong:2002rq}.
In order to make discussions clear under {\rm T}-duality transformations,
we summarize the scalar fields in the worldsheet sigma models
of various background geometries of five-branes and the labels of the spacetime coordinates $X^I$ in Table \ref{ST-WS}:
\begin{table}[ht]
\begin{center}
\slb{1}{\renewcommand{\arraystretch}{1.4}
\begin{tabular}{r||cccccc|cccc}
\hline
spacetime & 0 & 1 & 2 & 3 & 4 & 5 & 6 & 7 & 8 & 9 
\\ \hline \hline 
NS5-brane &
$\bigcirc$ & $\bigcirc$ & $\bigcirc$ & $\bigcirc$ & $\bigcirc$ & $\bigcirc$ & 
$r^1 = X^6$ & $r^3 = X^7$ & $r^2 = X^8$ & $\vartheta = X^9$
\\ 
KK-monopole & 
$\bigcirc$ & $\bigcirc$ & $\bigcirc$ & $\bigcirc$ & $\bigcirc$ & $\bigcirc$ & 
$r^1 = X^6$ & $r^3 = X^7$ & $r^2 = X^8$ & $\wt{\vartheta} = \wt{X}^9$
\\ 
exotic $5^2_2$-brane &
$\bigcirc$ & $\bigcirc$ & $\bigcirc$ & $\bigcirc$ & $\bigcirc$ & $\bigcirc$ & 
$r^1 = X^6$ & $r^3 = X^7$ & $y^2 = \wt{X}^8$ & $\wt{\vartheta} = \wt{X}^9$
\\ \hline
\end{tabular}
}
\end{center}
\caption{Correspondence between the worldsheet scalar fields and the spacetime coordinates.}
\label{ST-WS}
\end{table}

Finally, the FI parameters in the $\N=(4,4)$ theories are given by the pair of complex-valued variables such as
\begin{align}
s_a \ &= \ 
\frac{1}{\sqrt{2}} (s_{1,a} + \i s_{2,a})
\, , \ls
t_a \ = \ 
\frac{1}{\sqrt{2}} (t_{1,a} + \i \, t_{2,a})
\, ,
\end{align}
where $s_{i,a}$ and $t_{i,a}$ are real-valued.
In the same way as this, the other FI parameters $(s', t')$ in the main part of this paper are expanded.

\section{Remodeled GLSMs for defect five-branes}
\label{app-KKM2-H2}

In this appendix we briefly discuss the duality transformations \cite{Rocek:1991ps} of the Remodeled GLSM (\ref{X8X9-GLSM}) to other GLSMs for five-branes, i.e., the KK-monopole and the H-monopole.
Here we also perform the smearing procedure to the FI parameters $s_2$ (\ref{8-isometry}) and $s'_2$ (\ref{9-isometry}).
Then the target space geometries of the dualized NLSMs in the IR limit are deformed to the five-branes of codimension two.
We refer to these deformed five-branes as the defect KK-monopole and the defect NS5-brane, respectively \cite{Bergshoeff:2011se}.

\subsection{Remodeled GLSM for defect KK-monopole}
\label{app-KKM2}

We perform the duality transformation \cite{Rocek:1991ps} to the dynamical twisted chiral superfield $\Xi$ in the Remodeled GLSM for the exotic $5^2_2$-brane (\ref{X8X9-GLSM}). 
First, we convert the twisted F-term $\Xi \Sigma'$ in $\Scr{L}_{\text{G}}$ (\ref{X8X9-add}) to a D-term.
Next, we perform the conventional duality transformation \cite{Rocek:1991ps} to $\Xi$ in the similar way as (\ref{Theta2Gamma}).
Then $\Xi$ is dualized to a dynamical chiral superfield $\Psi'$:
\begin{gather}
- g^2 \Big\{ (\Xi + \ol{\Xi}) - \sqrt{2} (C + \ol{C}) \Big\}
- 2 V'
\ = \ 
\Psi' + \ol{\Psi}{}'
\, .
\end{gather}
Due to the previous relation (\ref{Psi2Xi}), 
we express the new duality transformation in the following form:
\begin{gather}
\Psi' + \ol{\Psi}{}' + 2 V'
\ = \ 
- g^2 (\Xi + \ol{\Xi})
+ \sqrt{2} \, g^2 (C + \ol{C})
\ = \ 
\Psi + \ol{\Psi}
\, . \label{dual-Psi'XiPsi}
\end{gather}
Notice that $\Psi$ is {\it not} the dynamical superfield
but merely a symbol which simplifies the expression of (\ref{dual-Psi'XiPsi}),
though originally $\Psi$ was the dynamical fields in the GLSM (\ref{GLSM-KKM}).
If the vector superfield $V'$ is turned off,
the symbol $\Psi$ corresponds to the dynamical field $\Psi'$.
Plugging (\ref{dual-Psi'XiPsi}) into (\ref{X8X9-GLSM}),
we obtain the Remodeled GLSM for the KK-monopole:
\begin{align}
\Scr{L}_{\text{KK2}}
\ &= \ 
\int \d^4 \theta \, \Big\{ 
\frac{1}{e^2} 
\Big( - \ol{\Sigma} \Sigma + \ol{\Phi} \Phi \Big)
+ \ol{Q} \, \e^{-2 V} Q
+ \ol{\wt{Q}} \, \e^{+2 V} \wt{Q}
\Big\}
\nn \\
\ & \ \ \ \ 
+ \int \d^4 \theta \, \Big\{
\frac{g^2}{2} 
\Big( \Gamma + \ol{\Gamma} + 2 V \Big)^2
+ \frac{1}{g^2} \ol{\Psi} \Psi
\Big\}
\nn \\
\ & \ \ \ \ 
+ \Big\{
\sqrt{2} \int \d^2 \theta \, \big( \wt{Q} \Phi Q + (s - \Psi) \, \Phi \big)
+ \text{(h.c.)} 
\Big\}
+ \Big\{ 
\sqrt{2} \int \d^2 \wt{\theta} \, t \, \Sigma
+ \text{(h.c.)}
\Big\}
\nn \\
\ & \ \ \ \ 
- \sqrt{2} \, \ve^{mn} \del_m (\vartheta A_{n})
- \sqrt{2} \, \ve^{mn} \del_m (y^2 A'_{n})
\nn \\
\ & \ \ \ \ 
+ \int \d^4 \theta \, \Big\{
\frac{1}{e'{}^2} \Big( - \ol{\Sigma}{}' \Sigma' + \ol{\Phi}{}' \Phi' \Big)
+ \ol{Q}{}' \, \e^{-2 V'} Q'
+ \ol{\wt{Q}}{}' \, \e^{+2 V'} \wt{Q}'
\Big\}
\nn \\
\ & \ \ \ \ 
+ \Big\{
\sqrt{2} \int \d^2 \theta \, \big( \wt{Q}' \Phi' Q'
+ (s' - \Gamma) \Phi' \big)
+ \text{(h.c.)} 
\Big\}
+ \Big\{ 
\sqrt{2} \int \d^2 \wt{\theta} \, t' \, \Sigma'
+ \text{(h.c.)}
\Big\}
\, . \label{X8X9-KKM}
\end{align}
As mentioned above, $\Psi$ is the functional of the superfields $\Psi'$ and $V'$ via (\ref{dual-Psi'XiPsi}).
This is slightly different from the GLSM for the KK-monopole (\ref{GLSM-KKM}) with $k = 1$.
This Remodeled GLSM possesses $\N=(2,2)$ supersymmetry rather than $\N=(4,4)$
caused by the same reason in the Remodeled GLSM (\ref{X8X9-GLSM}).

Following the same discussions in section \ref{sect-2-gauged}, 
we consider the IR limit of the Remodeled GLSM (\ref{X8X9-KKM}).
By a straightforward computation, we obtain 
\begin{align}
\Scr{L}_{\text{KK2b}}
\ &= \ 
- \frac{1}{2} \A
\Big\{ (\del_m r'{}^1)^2 + (\del_m r^3)^2 \Big\}
- \frac{1}{2} \B^{-1} 
\big( \del_m r'{}^2 - \C \, \del_m \wt{\vartheta} \big)^2
- \frac{1}{2} \B (\del_m \wt{\vartheta})^2
\nn \\
\ & \ \ \ \ 
- \sqrt{2} \, \ve^{mn} \del_m \big( (\vartheta - t_{2}) A_{n} \big)
- \sqrt{2} \, \ve^{mn} \del_m ((y^2 - t'_{2}) A'_{n})
\, . \label{X8X9-KKM-NLSM}
\end{align}
Here various functionals in (\ref{X8X9-KKM-NLSM}) are defined as follows:
\bsubeq \label{fns-KKM2}
\begin{align}
\A \ &= \ 
H_{\varrho} + \frac{1}{g^2} H'_{\varrho}
\, , \ls
\B \ = \ 
\frac{H_{\varrho}}{K_{\varrho}}
+ g^2 H'_{\varrho}
\, , \ls
\C \ = \ 
\frac{\omega_{\varrho}}{K_{\varrho}}
+ \omega'_{\varrho}
\, , \\
A_m \ &= \ 
- \frac{1}{\sqrt{2} \, g^2 H_{\varrho}}
\Big( \del_m \wt{\vartheta} 
- \omega_{\varrho} \, D_m r'{}^2 \Big) 
+ \frac{1}{\sqrt{2}} \del_m \wt{\vartheta}
\, , \ls
D_m r'{}^2 \ = \ 
\del_m r'{}^2 + \sqrt{2} A'_m
\, , \label{sol-A-X8X9-KKM} \\
A'_{m} 
\ &= \ 
\frac{1}{\sqrt{2}} \B^{-1}
\Big[ g^2 H'_{\varrho} 
\Big\{ \del_m r'{}^2 
- \frac{\omega_{\varrho}}{K_{\varrho}} 
\del_m \wt{\vartheta} 
\Big\}
+ \frac{H_{\varrho}}{K_{\varrho}} \, \omega'_{\varrho} \, \del_m \wt{\vartheta}
\Big]
\, . \label{sol-A'-X8X9-KKM} 
\end{align}
\esubeq
Here we have already performed the smearing procedure (\ref{8-isometry}) and (\ref{9-isometry}).
We fixed the gauge parameter $\alpha$ (or $\alpha'$) of the gauge field $A_m$ (or $A'_m$) to zero,
and rewrote $\gamma^4 = \wt{\vartheta}$ \cite{Tong:2002rq}.
Due to the duality transformation (\ref{dual-Psi'XiPsi}),
we replaced the scalar fields $(r^1, r^2)$ in the functionals $(H_{\varrho}, H'_{\varrho}, \omega_{\varrho}, \omega'_{\varrho})$ by $(r'{}^1, r'{}^2)$.

We extract the target space feature of the effective Lagrangian (\ref{X8X9-KKM-NLSM}).
The first line indicates the target space metric.
The second line denotes two types of dyonic modes \cite{Sen:1997zb}.
They are originated from the gauging of two isometries.
Compared with the string worldsheet sigma model (\ref{string-NLSM}),
we read off the explicit forms of the target space metric $G_{IJ}$ and the NS-NS B-field $B_{IJ}$:
\bsubeq \label{G-B-X8X9-KKM}
\begin{gather}
G_{66} \ = \ G_{77} \ = \ \A
\, , \ls
G_{88} \ = \ \B^{-1}
\, , \ls
G_{99} \ = \ 
\B + \C^2 \B^{-1}
\, , \\
G_{89} \ = \ 
- \C \B^{-1}
\, , \ls
B_{IJ} \ = \ 0
\, .
\end{gather}
\esubeq
This configuration does not satisfy the equations of motion in supergravity theories.
This is because the geometry is {\it not} Ricci-flat, even though the B-field is trivial.
In order to restore the Ricci-flatness on the geometry,
we should take the large $\wt{\cal R}_9$ limit, 
where the functionals $H'_{\varrho}$ and $\omega'_{\varrho}$ 
in (\ref{G-B-X8X9-KKM}) are proportional to the inverse of $\wt{\cal R}_9$.
As discussed before, the large $\wt{\cal R}_9$ limit implies that the radius of the physical coordinate $\wt{\vartheta} = \wt{X}^9$ is large.
In the same way, we take the small ${\cal R}_8$ limit, 
where ${\cal R}_8$ is the radius of the physical coordinate $r'{}^2 = X^8$.
In these limits,
the target space feature (\ref{G-B-X8X9-KKM}) is reduced to
\begin{gather}
G_{66} \ = \ G_{77} \ = \ H_{\varrho}
\, , \ls
G_{88} \ = \ \frac{K_{\varrho}}{H_{\varrho}}
\, , \ls
G_{99} \ = \ 
\frac{1}{H_{\varrho}}
\, , \ls
G_{89} \ = \ 
- \frac{\omega_{\varrho}}{H_{\varrho}}
\, . \label{G-B-dKKM}
\end{gather}
This is nothing but the background geometry of the defect KK-monopole.
We note that the defect KK-monopole is a five-brane of codimension two,
where the $X^8$-direction is compactified on a small radius from the standard geometry of the KK-monopole.
On the other hand, the Taub-NUT circle along the physical coordinate $\wt{X}^9$ is large.
Thus it is natural to take the above limits.

\subsection{Remodeled GLSM for defect NS5-brane}
\label{app-H2}

Here we study the dualization of the Remodeled GLSM for the defect KK-monopole (\ref{X8X9-KKM}) and perform the IR limit.
First we take the duality transformation \cite{Rocek:1991ps} to the dynamical chiral superfield $\Gamma$ in (\ref{X8X9-KKM}). 
Note that $\Gamma$ is coupled to $\Phi'$ in the F-term.
Then, in the similar way as (\ref{Psi2Xi}),
$\Gamma$ is dualized to a dynamical twisted chiral superfield $\Theta'$ \cite{Kimura:2013fda}:
\begin{align}
- g^2 \Big\{ (\Gamma + \ol{\Gamma}) + 2 V \Big\}
+ \sqrt{2} (C' + \ol{C}{}')
\ &= \ 
\Theta' + \ol{\Theta}{}'
\, . 
\end{align}
Here we used $\Phi' = \ol{D}{}_+ \ol{D}{}_- C'$.
Due to (\ref{Theta2Gamma}), this relation can be expressed as follows:
\begin{align}
\Theta' + \ol{\Theta}{}'
- \sqrt{2} (C' + \ol{C}{}')
\ &= \ 
- g^2 (\Gamma + \ol{\Gamma})
- 2 g^2 V 
\ = \ 
\Theta + \ol{\Theta}
\, . \label{dual-Theta'C'Gamma'}
\end{align}
We should notice that the twisted chiral superfield $\Theta$ 
is not a dynamical superfield but merely a symbol, or a functional of the other superfields.
Plugging (\ref{dual-Theta'C'Gamma'}) into (\ref{X8X9-KKM}),
we obtain another Remodeled GLSM\footnote{The duality transformation rule is close to the one in \cite{Kimura:2013fda}.}:
\begin{align}
\Scr{L}_{\text{H2}}
\ &= \ 
\int \d^4 \theta \, \Big\{ 
\frac{1}{e^2} 
\Big( - \ol{\Sigma} \Sigma + \ol{\Phi} \Phi \Big)
+ \ol{Q} \, \e^{-2 V} Q
+ \ol{\wt{Q}} \, \e^{+2 V} \wt{Q}
\Big\}
+ \int \d^4 \theta \, \frac{1}{g^2} 
\Big( - \ol{\Theta} \Theta + \ol{\Psi} \Psi \Big)
\nn \\
\ & \ \ \ \ 
+ \Big\{
\sqrt{2} \int \d^2 \theta \, \big( \wt{Q} \Phi Q + (s - \Psi) \, \Phi \big)
+ \text{(h.c.)} 
\Big\}
+ \Big\{ 
\sqrt{2} \int \d^2 \wt{\theta} \, \big( t - \Theta \big) \Sigma
+ \text{(h.c.)}
\Big\}
\nn \\
\ & \ \ \ \ 
+ \int \d^4 \theta \, \Big\{
\frac{1}{e'{}^2} \Big( - \ol{\Sigma}{}' \Sigma' + \ol{\Phi}{}' \Phi' \Big)
+ \ol{Q}{}' \, \e^{-2 V'} Q'
+ \ol{\wt{Q}}{}' \, \e^{+2 V'} \wt{Q}'
\Big\}
\nn \\
\ & \ \ \ \ 
+ \Big\{
\sqrt{2} \int \d^2 \theta \, \big( \wt{Q}' \Phi' Q' + s' \Phi' \big)
+ \text{(h.c.)} 
\Big\}
+ \Big\{ 
\sqrt{2} \int \d^2 \wt{\theta} \, t' \, \Sigma' + \text{(h.c.)}
\Big\}
\nn \\
\ & \ \ \ \ 
- \sqrt{2} \int \d^4 \theta \,
(\Gamma - \ol{\Gamma}) (C' - \ol{C}{}')
- \sqrt{2} \, \ve^{mn} \del_m (y^2 A'_{n})
\, . \label{X8X9-HM}
\end{align}
Here $\Psi$ and $\Theta$ are the functionals of the dynamical superfields $\Psi'$, $\Theta'$ and the vector multiplets via (\ref{dual-Psi'XiPsi}) and (\ref{dual-Theta'C'Gamma'}).
This is slightly different from the GLSM for the H-monopole (\ref{GLSM-HM}) with $k = 1$.
We note that the term $\Gamma - \ol{\Gamma}$ still remains even after the duality transformation (\ref{dual-Theta'C'Gamma'}). 
This implies that the imaginary part of $\Gamma$, i.e., the scalar field $\gamma^4$, behaves as the dual coordinate field,
as in the same way as the scalar field $r^2$ in the GLSM for the exotic $5^2_2$-brane (\ref{GLSM-522}).
The Remodeled GLSM (\ref{X8X9-HM}) has, 
as in the same way as (\ref{X8X9-GLSM}) and (\ref{X8X9-KKM}), 
only $\N=(2,2)$ supersymmetry, rather than $\N=(4,4)$ supersymmetry.

Following the discussions in section \ref{sect-2-gauged},
we consider the IR limit of the Remodeled GLSM (\ref{X8X9-HM}).
After a straightforward computation, we obtain
\begin{align}
\Scr{L}_{\text{H2b}}
\ &= \ 
- \frac{1}{2} \A \Big\{ (\del_m r'{}^1)^2 + (\del_m r'{}^3)^2 \Big\}
- \frac{1}{2} \B^{-1} (\del_m r'{}^2)^2 
- \half \big( \B + \C^2 \B^{-1} \big) (\del_m \wt{\vartheta})^2
\nn \\
\ & \ \ \ \ 
+ \C \B^{-1} (\del_m r'{}^2) (\del^m \wt{\vartheta}) 
+ \ve^{mn} (\del_m \wt{\vartheta}) (\del_n \vartheta')
- \sqrt{2} \, \ve^{mn} \del_m \big( (y^2 - t'_{2}) A'_{n} \big)
\, . \label{X8X9-HM-NLSM1}
\end{align}
Here we have already performed the smearing procedure (\ref{8-isometry}) and (\ref{9-isometry}).
The functionals in (\ref{X8X9-HM-NLSM1}) correspond to the ones (\ref{fns-KKM2}), whilst their variables $(r^3, \vartheta)$ are replaced by $(r'{}^3,\vartheta')$ via the duality transformation (\ref{dual-Theta'C'Gamma'}).
Since the dual field $\gamma^4 = \wt{\vartheta}$ still remains in the Lagrangian, we integrate it out. 
The solution is
\begin{align}
\del_m \wt{\vartheta} \ &= \ 
(\B + \C^2 \B^{-1})^{-1} 
\Big\{ \C \B^{-1} (\del_m r'{}^2) + \ve_{mn} (\del^n \vartheta') \Big\}
\, .
\end{align}
Substituting this into (\ref{X8X9-HM-NLSM1}), 
we obtain the final form of the NLSM:
\begin{align}
\Scr{L}_{\text{H2b}}
\ &= \ 
- \frac{1}{2} \A \Big\{ (\del_m r'{}^1)^2 + (\del_m r'{}^3)^2 \Big\}
- \frac{1}{2} \big( \B + \C^2 \B^{-1} \big)^{-1} 
\Big\{ (\del_m r'{}^2)^2 + (\del_m \vartheta')^2 \Big\}
\nn \\
\ & \ \ \ \ 
+ \C \B^{-1} \big( \B + \C^2 \B^{-1} \big)^{-1} 
\ve^{mn} (\del_m r'{}^2) (\del_n \vartheta') 
\nn \\
\ & \ \ \ \ 
- \sqrt{2} \, \ve^{mn} \del_m \big( (y^2 - t'_{2}) A'_{n} \big)
\, . \label{NLSM-X8X9-HM}
\end{align}
The first line in the right-hand side of (\ref{NLSM-X8X9-HM}) indicates the target space metric and the second line denotes the NS-NS B-field, 
while the third line gives rise to the dyonic mode.
Compared with the string worldsheet sigma model (\ref{string-NLSM}),
we can read the explicit forms of the target space variables:
\bsubeq \label{G-B-X8X9-HM}
\begin{gather}
G_{66} \ = \ 
G_{77} \ = \ 
\A
\, , \ls
G_{88} \ = \ 
G_{99} \ = \ 
\big( \B + \C^2 \B^{-1} \big)^{-1}
\, , \\
B_{89} \ = \ 
\C \B^{-1}
\big( \B + \C^2 \B^{-1} \big)^{-1}
\, .
\end{gather}
\esubeq
This configuration does not satisfy the equations of motion in supergravity theories.
In order to satisfy these equations,
we should take the large $\wt{\cal R}_9$ limit, 
where the functionals $H'_{\varrho}$ and $\omega'_{\varrho}$ 
in (\ref{G-B-X8X9-HM}) are proportional to the inverse of $\wt{\cal R}_9$.
As discussed before, the large $\wt{\cal R}_9$ limit implies that the radius of the dual coordinate $\wt{\vartheta} = \wt{X}^9$ is large.
This also denotes that the radius ${\cal R}_9$ of the physical coordinate $\vartheta' = X^9$ is small, 
because the two radii are related to each other via ${\cal R}_9 =
\alpha'/\wt{\cal R}_9$
where $\alpha'$ is the string Regge slope parameter.
In the same way, we take the small ${\cal R}_8$ limit, 
where ${\cal R}_8$ is the radius of the physical coordinate $r'{}^2 = X^8$.
In these limits,
the target space feature (\ref{G-B-X8X9-HM}) is reduced to
\begin{gather}
G_{66} \ = \ 
G_{77} \ = \ 
G_{88} \ = \ 
G_{99} \ = \ 
H_{\varrho}
\, , \ls
B_{89} \ = \ 
\omega_{\varrho}
\, . \label{G-B-dHM}
\end{gather}
This is nothing but the background geometry of the defect NS5-brane.
We note that the defect NS5-brane is a five-brane of codimension two,
where both the $X^8$- and $X^9$-directions are compactified on small radii from the standard geometry of the H-monopole.
Thus it is natural to take the above limits.

\end{appendix}

}
\end{document}